\documentclass{aastex6}
\usepackage{epsf}
\usepackage{epsfig}
\usepackage{amsmath,amsfonts,amsthm, amssymb,latexsym}
\usepackage{enumerate}
\usepackage{graphicx}% Include figure files
\usepackage{dcolumn}% Align table columns on decimal point
\usepackage{bm}% bold math
\usepackage{amsmath}	% Advanced maths commands
\usepackage{amssymb}	% Extra maths symbols
\usepackage{color}

\def\be{\begin{equation}}
\def\ee{\end{equation}}

\begin{document}

\title{Many-body forces in magnetic neutron stars}
\date{\today}
\author{R.O. Gomes}
\affiliation{Astronomy Department, Universidade Federal do Rio Grande do Sul, Porto Alegre, Brazil}
\affiliation{Frankfurt Institute for Advanced Studies,
Frankfurt am Main, Germany}
\email{rosana.gomes@ufrgs.br}
\author{B. Franzon}
\affiliation{Frankfurt Institute for Advanced Studies,
Frankfurt am Main, Germany}
\author{V. Dexheimer}
\affiliation{Department of Physics, Kent State University, Kent OH, USA}
\author{S. Schramm}
\affiliation{Frankfurt Institute for Advanced Studies,
Frankfurt am Main, Germany}

\begin{abstract}
In this work, we study in detail the effects of many-body forces on the equation of state and the structure of magnetic neutron stars. 
The stellar matter is described within a relativistic mean field formalism that takes into account many-body forces by means of a non-linear meson field dependence on the nuclear interaction coupling constants. We assume that matter is at zero temperature, charge neutral, in beta-equilibrium, and populated by the baryon octet, electrons, and muons. In order to study the effects of different degrees of stiffness in the equation of state, we explore the parameter space of the model, which reproduces nuclear matter properties at saturation, as well as massive neutron stars. 
Magnetic field effects are introduced both in the equation of state and in the macroscopic structure of stars by the self-consistent solution of the Einstein-Maxwell equations. In addition, effects of poloidal magnetic fields on the global properties of stars, as well as density and magnetic field profiles are investigated. We find that not only different macroscopic  magnetic field distributions, but also different parameterizations of the model for a fixed magnetic field distribution impact the gravitational mass, deformation and internal density profiles of stars.
Finally, we also show that strong magnetic fields affect significantly the particle populations of stars.%, and that including magnetic effects on the structure of stars have a strong impact on their global properties.
\end{abstract}

\maketitle

\section{Introduction}

Neutron stars are one of the possible outcomes of the evolution of heavy stars. The transition from a main sequence star to a compact remnant happens through gravitational collapse, after depletion of  nuclear fuel. During the collapse, matter in the core of these stars is compressed beyond the point of neutron drip, reaching densities of many times nuclear saturation density. The collapse is then stopped by repulsive nuclear interactions that are dominant in this regime. Because of angular momentum and magnetic flux conservation, the rotation rates and magnetic fields of these stars are exceptionally amplified, during the collapse reaching the typical values of $P \sim 1 \, \mathrm{s} $ and $B_{s} \sim 10^{12}\,\mathrm{G}$, respectively. 
Considered the densest stars known, neutron stars have about the mass of the Sun compressed in a small radius of roughly $10\,\mathrm{km}$, harboring ultra-dense nuclear matter in their interiors. These objects provide a unique environment for investigating fundamental questions in physics and astrophysics, including matter under extreme conditions, such as the effects of gravity, magnetic fields and nuclear interactions in the strong regime. 

Certain classes of neutrons stars known as Soft Gamma Repeaters (SGRs) and Anomalous X-ray Pulsars (AXPs) present surface magnetic fields of the order on $B_{s} \sim 10^{15}\,\mathrm{G}$, which is much higher than the ones present in normal pulsars. These objects are called \emph{magnetars}. Although the internal magnetic fields cannot be constrained by observations, the virial theorem predicts that the internal magnetic fields of magnetars can reach up to $B_c \sim 10^{18}-10^{20}\,\mathrm{G}$ in the stellar center  \cite{Lai1991,Cardall:2000bs,Ferrer:2010wz}. It is estimated that approximately $10\%$ of neutron stars are magnetars \cite{Kouveliotou:1998ze,magcatalog} \footnote{ http://www.physics.mcgill.ca/\tiny{$\sim$}\normalsize{  pulsar/magnetar/main.html}}.

The origin of such high magnetic fields is still under debate. Note, for example, that the magnetic flux conservation during the gravitational collapse does not suffice as explanation of the extreme magnetic fields present in magnetars. In this case, a $1.4\,\mathrm{M_{\odot}}$ star would need to have a radius smaller than its Schwarzschild  radius in order to generate a magnetic field of $B \sim 10^{15}\,\mathrm{G}$ \cite{Tatsumi:1999ab}. 
The \emph{magnetohydrodynamic dynamo mechanism} (MHD) is currently the most accepted theory to describe the surface magnetic fields in magnetars. This theory, developed by Duncan \& Thompson in the 90s \cite{Duncan:1992hi,Thompson:1993hn}, is based on the amplification of the stellar magnetic fields through the combination of rapid rotation and convective processes in the plasma during the proto-neutron star phase. Hence, such an effective dynamo mechanism can build up strong magnetic fields compatible with the ones present in magnetars. After about $\sim 10 \, \mathrm{s}$ (end of proto-neutron star phase), thermal effects become irrelevant, stopping convection processes, and leaving behind highly magnetized neutron stars.
The MHD mechanism successfully explains several phenomena that occur on the surface of magnetars, such as their steady X-ray emission, gamma-ray explosions, and giant flare events \cite{Thompson:1995gw,Thompson:1996pe}. 

In order to consistently describe magnetars in a general relativistic framework, it is necessary to take magnetic fields into account to model both the equation of state of matter inside stars and their global structure. 
The first studies on strong magnetic fields effects in a low-density Fermi gas were performed by Canuto  \cite{Chiu:1968zz,Canuto:1969ct,Canuto:1969cs,Canuto:1969cn}, 
followed by calculations using more realistic hadronic models with magnetic field effects  \cite{Chakrabarty:1997ef,Broderick:2000pe}. 
Over the years, it has been shown that the presence of strong magnetic fields significantly affects the energy levels of charged particles due to  Landau quantization. These effects have been taken into account in the equation of state (EoS) of models in order to describe hyperon stars  \cite{Chakrabarty:1997ef,Broderick:2000pe,Broderick:2001qw,Sinha:2010fm,Lopes:2012nf,Casali:2013jka,Gomes:2014dka,Gao:2015jha}, quark stars  \cite{PerezMartinez:2007kw,Orsaria:2010xx,Dexheimer:2012mk,Denke:2013gha,Isayev:2015rda,Felipe:2010vr,Paulucci:2010uj},  
hybrid stars \cite{Rabhi:2009ih,Dexheimer:2011pz,Dexheimer:2012qk}, and  stars  with meson condensates \cite{Schramm:2015lga}. 
Effects of the anomalous magnetic moment were also investigated  through the corresponding coupling of the baryons to the electromagnetic field tensor (see Refs. \cite{Canuto:1969cn,Broderick:2000pe,Broderick:2001qw,Strickland:2012vu}).
Moreover, Landau quantization effects on the EoS gives rise to an anisotropy in the matter energy-momentum tensor components \cite{PerezMartinez:2007kw,Strickland:2012vu}, indicating that deformation effects can be important in the microscopic description of these objects and, therefore, that spherical symmetry is not the best approximation to describe the macroscopic structure of magnetized neutron stars. 

The inclusion of magnetic field effects in the calculation of the stellar macroscopic structure involves solving the coupled  Einstein-Maxwell equations using a metric able to describe deformed objects, i.e., at least a two-dimensional metric.
%Solving such a system of equations is a challenging task from the numerical and analytical point of view.  
Simplified solutions for the problem were proposed by \cite{Schramm:2013ipa,Paret:2014sba,Zubairi:2015yda}, where a perturbation of the metric was carried out. 
 %magnetic fields %for an object with quasi-spherical symmetry  . 
Numerically, the formalism developed by Bonazzola et al. \cite{Bonazzola:1993zz}, implemented in the LORENE library, takes into account both rotation and magnetic fields in a full calculation of the stellar structure of neutrons stars.
This formalism was initially applied to a one-parameter equation of state \cite{Bocquet:1995je,Cardall:2000bs}. Only recently, self-consistent calculations including magnetic fields both in the equation of state and stellar structure were implemented to describe quark stars in Ref. \cite{Chatterjee:2014qsa} and hybrid stars in Ref. \cite{Franzon:2015sya}.

Although recent self-consistent calculations for magnetic neutron stars have shown that the Landau quantization effects barely impact the global properties of such objects \cite{Chatterjee:2014qsa,Franzon:2015sya}, as a side product of this work, we cross-check these results  for the case of hyperon stars complementing previous results for quark and hybrid stars. 
Another important analysis, which is lacking in the literature, regards the  impact of the EoS stiffness on strongly magnetized stars. Calculations of the structure of magnetic neutron stars were performed using  different hadronic models \cite{Bocquet:1995je,Cardall:2000bs,Dexheimer:2016yqu},  but the focus of the analysis has always been on the impact of different magnetic field configurations on the global properties of stars using one specific model and parametrization. 

The  determination  of  the  nuclear matter EoS at high densities is still an open question and also one of the main  goals  of current  nuclear  astrophysics research. Since current lattice QCD calculations cannot reach the regime of high densities due to the highly oscillatory behavior in the functional integral, it is not possible to describe the EoS of dense, strongly interacting matter on a fundamental level. However, assuming that only baryonic degrees of freedom are relevant for the energy scales present in neutron stars, the nuclear interaction can be reasonably approximated
by effective relativistic mean field hadronic models. In such approaches, the baryon-baryon interaction is described by the exchange of scalar and vector mesons, which simulate the attractive and repulsive features of the nuclear interaction. In the past, different couplings where suggested, including models with non-linear contributions of the meson fields \cite{Boguta:1977xi,Sugahara:1993wz,Toki1995,ToddRutel:2005fa,Kumar:2006ij}, density dependent couplings \cite{Typel:1999yq}, meson field dependence \cite{Zimanyi:1990np,Taurines:2000zb,Dexheimer:2007mt,Gomes:2014aka}, among others. In particular, the many-body forces model (MBF model) \cite{Taurines:2000zb,Dexheimer:2007mt,Gomes:2014aka} introduces a field dependence of the couplings, making them  indirectly density dependent. The non-linear terms that arise from the coupling expansion are an important feature of the MBF model, as these contributions can be interpreted as contributions from many-body forces due to meson-meson interactions.

\begin{table*}
  \caption{\label{Hmodels} Basic nuclear properties at saturation and couplings, for different parameterizations of the MBF model. The columns are: the many-body forces parameter $\zeta$, nucleon efffective mass $m^*_n$, compressibility modulus $K_0$, and the mesons couplings ($g_{\sigma N}$, $g_{\omega N}$, $g_{\varrho N}$, $g_{\delta N}$). }

\begin{center}
\begin{tabular}{ccccccc}
 \hline
$\zeta$ & $m^*_n/m_n$ &  $K_0$~(MeV) & $(g_{\sigma N}/m_{\sigma})^2$ & $(g_{\omega N}/m_{\omega})^2$ & $(g_{\varrho N}/m_{\varrho})^2$ & $(g_{\delta N}/m_{\delta})^2$   \\
  \hline \hline
    %&  &  &  &  \\
 0.040 & 0.66 & 297  & 14.51  & 8.74 & 4.466 & 0.383 \tabularnewline
 0.059 & 0.70 & 253  & 13.44  & 7.55 & 5.571 & 1.820 \tabularnewline
 0.085 & 0.74 & 225  & 12.21  & 6.37 & 6.467 & 3.103 \tabularnewline 
 0.129 & 0.78 & 211  & 10.84  & 5.16 & 7.057 & 4.038 \tabularnewline 
  \hline\hline
  \end{tabular}
\end{center}
\end{table*}

The observation of massive neutron stars \cite{Demorest2010,Antoniadis2013} indicates that the EoS of nuclear matter must be very stiff in the regime of high densities and low temperatures. 
The degree of stiffness in the nuclear matter EoS is directly related to the repulsive interaction among particles at high densities, as well as to the particle content in the core of the stars. 
In particular, it has been extensively discussed in the literature whether or not exotic degrees of freedom  might populate the core of neutron stars. On one hand, it is more energetically favorable for the system to populate new degrees of freedom, such as hyperons
\cite{Ishizuka:2008gr,Dexheimer:2008ax,Bednarek:2011gd,Gomes:2014aka,Oertel:2014qza,Lonardoni:2014bwa,Burgio:2015zka,Fukukawa:2015iba,Lonardoni:2015iba,Maslov:2015msa,Chatterjee:2015pua,Vidana:2015rsa,Yamamoto:2015lwa,Vidana:2015rsa,Torres:2016ydl,Biswal:2016zcg,Tolos:2016hhl,Mishra:2016qhw}, delta isobars \cite{Schurhoff:2010ph,Fong:2010zz,Drago:2013fsa,Drago:2015cea,Cai:2015hya,Zhu:2016mtc}, and meson condensates \cite{Takahashi:2007qu,Ohnishi:2008ng,Ellis:1995kz,Menezes:2005ic,Mishra:2009bp,Alford:2009jm,Fernandez:2010zzc,Mesquita:2010zzb,Lim:2013tqa,Muto:2015sgx}, in order to lower its Fermi energy (starting at about two times the saturation density). On the other hand, the EoS softening due to the appearance of exotica might turn some nuclear models incompatible with observational data, in particular with the recently measured massive neutron stars. One possible way to overcome this puzzle is the introduction %of strange mesons, which introduces 
of an extra repulsion in the YY interaction \cite{Schaffner:1995th,Bombaci:2016xzl}, allowing  models with hyperons to be able to reproduce massive stars \cite{Dexheimer:2008ax,Bednarek:2011gd,Weissenborn:2011kb,Lopes:2013cpa,Banik:2014qja,Gomes:2014aka,Bhowmick:2014pma,vanDalen:2014mqa,Gusakov:2014ota,Yamamoto:2014jga}.
Another possible solution is the introduction of a deconfinement phase transition at high densities \cite{Bombaci:2016xzl}, with a stiff EOS for quark matter, usually associated with quark vector interactions (see Ref. \cite{Klahn:2013kga} and references therein).

In this work, we investigate the impact of different parameterizations of the MBF model and different magnetic field configurations on the global properties and  on the density and magnetic profiles of magnetic neutron stars. First, we calculate an error estimate for using  spherical TOV solutions for magnetic stars (instead of a two-dimensional deformed solution).  
In order to uniquely identify the dependence of global stellar properties on the EoS, we fix the baryonic mass of the stars. 
%and the magnetic field configuration to the maximum value that leads to numerically stable results. 
Then, we study different stellar properties such as gravitational mass, deformation, density and magnetic field profiles, and particle population. 

This paper is organized as follows: in Section II the MBF model with the inclusion of magnetic fields is presented; in Section III we discuss the main features of the Einstein-Maxwell equations; Section IV estimates the error in neglecting stellar deformation when  describing magnetic neutron stars; in Section V we show the impact of different degrees of EoS stiffness and different magnetic field distributions on the global properties and internal magnetic and density profiles of  
a fixed baryon mass star; Section VI is dedicated to discuss the particle populations of the stars and, finally, in Section VII we present our conclusions.

\section{The Many-body forces formalism}

In this section, we present the model used for describing microscopic matter in this work.
We also describe how magnetic fields modify the model through the introduction of Landau quantization.

The Lagrangian density of the MBF model presented in Ref. \cite{Gomes:2014aka} includes for the first time a coupling dependence on all scalar fields. It reads:

\small
\begin{equation}\begin{split}\label{lagrangian}
\mathcal{L}&= \underset{b}{\sum}\overline{\psi}_{b}\left[\gamma_{\mu}\left(i\partial^{\mu} -g_{\omega b}\omega^{\mu} -g_{\phi b}\phi^{\mu}
-g_{\varrho b}\mathbf{\textrm{\ensuremath{I_{3b}}\ensuremath{\varrho_3^{\mu}}}}\right)
-m^*_{b \zeta}\right]\psi_{b}
 +\left(\frac{1}{2}\partial_{\mu}\sigma\partial^{\mu}\sigma-m_{\sigma}^{2}\sigma^{2}\right)
+\frac{1}{2}\left(-\frac{1}{2}\omega_{\mu\nu}\omega^{\mu\nu}+m_{\omega}^{2}\omega_{\mu}\omega^{\mu}\right)
\\&+\frac{1}{2}\left(-\frac{1}{2}\boldsymbol{\varrho_{\mu\nu}.\varrho^{\mu\nu}}+m_{\varrho}^{2}\boldsymbol{\varrho_{\mu}.\varrho^{\mu}}\right)
+\left(\frac{1}{2}\partial_{\mu}\boldsymbol{\delta.}\partial^{\mu}\boldsymbol{\delta}-m_{\delta}^{2}\boldsymbol{\delta}^{2}\right)
+\frac{1}{2}\left(-\frac{1}{2}\phi_{\mu\nu}\phi^{\mu\nu}+m_{\phi}^{2}\phi_{\mu}\phi^{\mu}\right) 
+\underset{l}{\sum}
\overline{\psi}_{l}\gamma_{\mu}\left(i\partial^{\mu}
-m_{l}\right)\psi_{l}
\\& +\underset{b}{\sum}\overline{\psi}_{b}q_{e,b}A^{\mu}\psi_{b}
 +\underset{l}{\sum}\overline{\psi}_{l}q_{e,l}A^{\mu}\psi_{l}.
\end{split}\end{equation}
\normalsize

The subscripts $b$ and $l$ correspond to baryons ($p^+$, $n^0$, $\Lambda^0$, $\Sigma^+$, $\Sigma^0$, $\Sigma^-$, $\Xi^0$, $\Xi^-$) and leptons ($e^-$, $\mu^-$) degrees of freedom, respectively.
The first line and the second term in the fourth line of Eq. (\ref{lagrangian}) represent the Dirac Lagrangians for baryons and leptons.
The electromagnetic interaction is introduced by the coupling to the photon field $A^{\mu}$ field, in the last line, and other terms represent the Lagrangian densities of scalar mesons ($\sigma$, $\delta$) and vector mesons ($\omega$, $\varrho$, $\phi$).
The meson-baryon coupling appears in the first term for the vector mesons ($g_{\omega b}$, $g_{\varrho b}$, $g_{\phi b}$) and the scalar ones are contained in the baryon effective masses ($m^ {*}_{\zeta b}$). 
The introduction of the $\delta$ and $\varrho$ isovector fields 
is important to extrapolate the model to isospin asymmetric systems, such as neutron stars, and the $\phi$ vector meson adds a repulsive component in the YY interaction.

The effects of the many-body forces contribution is introduced in the effective couplings of the scalar mesons: 
\begin{equation}
g^{*}_{i b} =\left(1+ \frac{g_{\sigma b}\sigma+ g_{\delta b}I_{3b}\delta_{3}}{\zeta m_{b}}  \right)^{-\zeta} g_{i b}, 
\label{geff}
\end{equation}
for $i=\sigma,\,\delta$ and, consequently, in the baryon masses as:
\begin{equation}
m_b^* = m_b - \left(  g^{*}_{\sigma b}\sigma + g^{*}_{\delta b}I_{3b}\delta_{3} \right).
\label{meff}
\end{equation}
If we were to expand the scalar couplings around the $\zeta$ parameter, we would generate nonlinear contributions from self and crossed terms of the scalar fields ($\sigma$, $\delta$).
These couplings simulate the effects of many-body forces in the nuclear interaction, which are controlled by the $\zeta$ parameter. 
In this way, each value of the parameter generates a different EoS and, hence, a different set of nuclear saturation properties.

In this work, we vary the $\zeta$ values obeying the following constraints for symmetric matter: binding energy per nucleon $B/A = -15.75\,\mathrm{MeV}$ and saturation density $\rho_0 = 0.15\,\mathrm{fm^{-3}}$.
The values of the nuclear asymmetric matter properties at saturation are fixed as: symmetry energy $a_{sym}=32\,\mathrm{MeV}$ and its slope $L_0=97\,\mathrm{MeV}$. We choose the lowest value of the slope which is common to all parametrizations of the MBF model, corresponding to the 
lowest value for the parametrization $\zeta = 0.040$ 
%. Other parametrizations of the model allow for lower values of this quantity 
\cite{Gomes:2014aka}. The corresponding remaining nuclear properties at saturation, associated to the sets of parameters used in this work (in agreement with experimental data) are collected in Table \ref{Hmodels}.

%and chemical potentials are expressed in the effective masses and the effective chemical potential:
%\begin{equation}\label{mu_eff}
% \mu^*_{b_i}= e_{F,b} + g_{\omega b}\omega +g_{\varrho b}  I_{3b} \varrho_{3} + g_{\phi b}\phi,
%\end{equation}
%where $k_{f_b}$ and $m_b$ correspond to the fermi momenta and the masses of the baryons.

When the hyperon degrees of freedom are taken into account throughout this paper, we describe their vector couplings using the SU(6) spin-flavor symmetry \cite{Dover:1985ba,Schaffner:1993qj}. We fit the hyperon-sigma coupling in order to reproduce the following values of the hyperon potential depths \cite{SchaffnerBielich:2000wj}:
$U_{\Lambda}^N =-28\, \mathrm{MeV}$, $U_{\Sigma}^N=+ 30\, \mathrm{MeV}$, and $U_{\Xi}^N= -18\, \mathrm{MeV}$. 
However, it is important to mention that only two ($\zeta = 0.040,\,0.059$) of the parametrizations  in Table \ref{Hmodels} describe hyperon stars in agreement with the observational data for massive stars \cite{Gomes:2014aka}. Also, although some parametrizations of the MBF model are very stiff, note that the formalism is relativistically invariant, consequently, obeys causality.

The introduction of magnetic fields alters the energy levels of the charged particles, which become Landau quantized as follows:
\begin{equation}\label{landau_levels}
e_{F}=\left\{ \begin{array}{c}
\sqrt{\left(m_{i}\right)^{2}+k_{F}^{2}},\qquad \qquad \quad   q=0\\
\sqrt{\left(m_{i}\right)^{2}+k_{F,z}^{2}+2|q|B\nu}, \quad q\neq 0
\end{array}\right.,
\end{equation}
for uncharged and charged fermions, respectively, with $m_{i} = m_i^{*}$ in the case of baryons. The Landau quantum number $\nu$ is given in terms of the orbital and spin quantum numbers as:
\begin{equation}
\nu \equiv  l + \frac{1}{2} -  \frac{s}{2} \frac{q}{|q|},
\end{equation}
running within the range $0 \leq \nu \leq \nu_{max}$, with denoting the highest Landau orbit with non-vanishing particle occupation at zero temperature (see Ref. \cite{Strickland:2012vu} for more details).

\section{The Einstein-Maxwell solutions}

In what follows, we introduce the formalism used to calculate the macroscopic structure of magnetic neutron stars. %and then apply it to the different parameterizations of the MBF model. For this purpose, we calculate for one parametrization a EoS for different magnetic field distributions.
The spacetime in general relativity is described by the Einstein equations (EE):
\be
R_{\mu\nu} - \frac{1}{2}R g_{\mu\nu} = 8\pi G T_{\mu\nu},
\label{eefinal} 
\ee
with $R_{\mu\nu}$ being the Ricci tensor, $R$ the Ricci scalar, $T_{\mu\nu}$ the energy-momentum tensor of matter and electromagnetic fields, and $G$ the Newton's gravitational constant.

With the assumption of a stationary, axi-symmetric
spacetime, and Maximum-Slice-Quasi-Isotropic coordinates (MSQI), the  line element in the 3+1 decomposition of space-time can be cast in the form:
\be
ds^2 = -N^{2}dt^{2} + A^2(dr^2 + r^2 d\theta^2) + B^{2}r^{2}\sin^{2}\theta( d\phi - \omega dt)^{2},
\label{metric}
\ee 
with $N(r,\theta)$, $A(r,\theta)$, $B(r,\theta)$, and $\omega(r, \theta)$ being functions only of the coordinates $(r, \theta)$. In this case, the final system of equations for each metric potential is given by:
\be 
\Delta_{2} [(NB-1) r \sin \theta] = 8\pi NA^2 B r \sin \theta (S^{r}_{r} + S^{\theta}_{\theta}),
\label{Bfinal}
\ee
\be 
\Delta_{2} [{\rm{ln}} A + \nu] = 8\pi A^2 S^{\phi}_{\phi} + \frac{3 B^2 r^2 \sin^2 \theta}{4 N^2} \partial \omega \partial \omega - \partial \nu \partial \nu,
\label{Afinal}
\ee
\be 
\Delta_{3} \nu = 4\pi A^2 (E + S) + \frac{B^2 r^2 \sin^2 \theta}{2N^2} \partial \omega \partial \omega - \partial \nu \partial( \nu + {\rm{ln}} B),
\label{Nfinal}
\ee
and 
\begin{align}
&\left[ \Delta_{3} - \frac{1}{r^2 \sin^2 \theta} \right] (\omega r \sin \theta) = \nonumber -16\pi \frac{NA^2}{B^2} \frac{p_{\phi}}{r \sin \theta} 
 + r \sin \theta  \partial \omega \partial(\nu - 3 {\rm{ln}} B),
\label{omegafinal}
\end{align}
where the short notation was introduced:
\begin{align}
& \Delta_{2} = \frac{ \partial^2}{\partial r^2} + \frac{1}{r}\frac{ \partial}{\partial r} + \frac{1}{r^2}\frac{ \partial^2}{\partial \theta^2},\\
& \Delta_{3} = \frac{ \partial^2}{\partial r^2} + \frac{2}{r}\frac{ \partial}{\partial r} + \frac{1}{r^2}\frac{ \partial^2}{\partial \theta^2} + \frac{1}{r^2 \tan \theta}\frac{ \partial}{\partial \theta}, \\
& \nu = {\rm{ln}} N,
\end{align}
with the total energy of the system  $E = E^{PF} + E^{EM}$, the total momentum density flux $J_{\phi} = J^{PF}_{\phi} +  J^{EM}_{\phi}$ and the total stress tensor $S = S^{PF} + S^{EM}$, where $PF$ and $EM$ stand for the perfect fluid and the electromagnetic field contributions, respectively. In addition, in the final system of gravitational field equations  Eqs.~\eqref{Bfinal}-\eqref{omegafinal}, terms such as $\partial \omega \partial \omega$ are defined as:
\be 
\partial \omega \partial \omega := \frac{\partial \omega}{\partial \omega}\frac{\partial \omega}{\partial r} + \frac{1}{r^2}\frac{\partial \omega}{\partial \theta}\frac{\partial \omega}{\partial \theta}.
\label{notation}
\ee

%To close, the equation of motion, $\nabla_{\mu} T^{\mu\nu} =0$, for the matter in the MSQI coordinates is given by:

%\be
%\frac{N}{\sqrt{\gamma}}\frac{\partial}{\partial x^{j}} \left( \sqrt{\gamma} S^{i}_{i}\right) - \frac{N}{2}\frac{\partial \gamma_{jk}}{\partial x^{i}}S ^{jk} + S^{jk}\frac{\partial N}{\partial x^{j}} + E \frac{\partial N}{\partial x^{i}} + J_{\phi} \frac{\partial \omega} {\partial x^{i}} = 0,
%\label{equationofmotionfinal}
%\ee 
%where $x^{i}$ stands for $r$ and $\theta$. In the next pages this equation of motion when applied to a perfect fluid with and without a magnetic field. 

The Faraday tensor $F_{\mu\nu}$ is derived from the magnetic 4-vector potential $A_{\mu}$ as: 
\be 
F_{\mu\nu} = A_{\nu, \mu} - A_{\mu,\nu}, 
\ee
such that the Maxwell equation:
\be
F_{\alpha\beta;\gamma} + F_{\beta\gamma;\alpha} + F_{\gamma\alpha;\beta} = 0,
\ee
is automatically satisfied.  The remaining Maxwell equations can be expressed in terms of the two non-vanishing components of the magnetic vector potential $A_{\mu}$ as:
\be 
F^{\alpha\beta}_{;\beta} = 4 \pi j^{\alpha}.
\ee
For example, the Maxwell-Gauss equation for $A_{t}$ reads:
\begin{align} 
\Delta_{3}A_{t} = &- \mu_{0} A^2 (g_{tt}j^{t} + g_{t\phi}j^{\phi}) \nonumber - \frac{B^2}{N^2}\omega r^2 \sin^2 \theta \partial A_{t} \partial N^{\phi} \nonumber -\left(1+\frac{B^2}{N^2} r^2 \sin^2 \theta \omega^2 \right) \partial A_{\phi}\partial \omega \nonumber  \\
&-(\partial A_{t} + 2 \omega \partial A_{\phi}) \partial ({\rm{ln}}B - \nu) -2\frac{\omega}{r} \left( \frac{\partial A_{\phi}}{\partial r} + \frac{1}{r \tan\theta} \frac{\partial A_{\phi}}{\partial r}\right),
\label{maxwellgauss}
\end{align}
and from the Maxwell-Ampere equation, we have an equation for $A_{\phi}$:
\begin{align}
\left[\Delta_{3} -\frac{1}{r^2 \sin^2 \theta}\right] & \left( \frac{A_{\phi}}{r \sin \theta}\right)\nonumber  =  - \mu_{0} A^2 B^2 ( j^{\phi} - \omega j^{t})r \sin \theta \nonumber \\
& - \frac{B^2}{N^2} r \sin \theta \partial \omega (\partial A_{t}+\omega \partial A_{\phi})
+ \frac{1}{r} \partial A_{\phi} \partial ({\rm{ln}}B - \nu).
\label{maxwellampere}
\end{align}

In this approach, the equation of motion reads:
\be 
H\left(r, \theta \right) + \nu\left(r, \theta \right) + M\left(r, \theta \right) = const,
\ee
with $H(r,\theta)$ being the heat function defined in terms of the
baryon number density $n$ and the magnetic potential $M(r,\theta)$ is given by: 
\be H =
\int^{n}_{0}\frac{1}{e(n_{1})+p(n_{1})}\frac{d p}{dn}(n_{1})dn_{1},
\label{heat}
\ee
\be M \left(r, \theta \right) = M \left(
A_{\phi} \left(r, \theta \right) \right) \equiv -
\int^{0}_{A_{\phi}\left(r, \theta \right)} f\left(x\right)
\mathrm{d}x, 
\ee
respectively, with  $f(x)$ being the current function, which we choose to be constant in this work.  According to Ref.~\cite{Bocquet:1995je}, other choices for $f(x)$ are possible, but they do not change the results qualitatively in the polar direction. The quantitative differences are explored by using different electric amplitudes $j_0$.
From the current function and the equation of state, one obtains the electric current: 
\be 
j^{\phi} - \Omega j^{t} = (e+p)\,f(x),
\label{currentEOS}
\ee
where $e$ and $p$ are the energy density and pressure of matter, respectively.

The stress-energy tensor of the magnetic field is calculated using the standard expression:
\be 
T^{EM}_{\alpha\beta} = \frac{1}{4 \pi} \left( F_{\alpha \mu} F^{\mu}_{\beta} - \frac{1}{4} F_{\mu\nu} F^{\mu\nu} \mathrm{g}_{\alpha\beta} \right),
\ee
from which one can obtain the sources of the gravitational fields. The electromagnetic contribution to the total energy and to the momentum density of the system are, respectively:
\be 
E^{EM} = \frac{1}{2 \mu_{0}} B^{i}B_{i},
\label{EMenergydensity}
\ee
%while the EM contribution to the momentum density can be written as
\be 
J^{EM}_{\phi} = \frac{1}{\mu_{0}} A^2 (B^r E^{\theta} - E^r B^{\theta}).
\label{EMmomentumdensity}
\ee
The stress 3-tensor components are given by:
\begin{align}
&S^{EM r}_{\;\;\; r} = \frac{1}{2 \mu_{0}}( E^{\theta}E_{\theta} - E^r E_{r} + B^{\theta}B_{\theta} - B^r B_{r}),\\
&S^{EM \theta}_{\;\;\; \theta} = \frac{1}{2 \mu_{0}}( E^r E_{r} - E^{\theta}E_{\theta}  +B^r B_{r}-  B^{\theta}B_{\theta}),\\
&S^{EM \phi}_{\;\;\; \phi} = \frac{1}{2 \mu_{0}} (E^{i}E_{i} + B^{i}B_{i}),
\label{EMstress}
\end{align} 
being the electric field components, as measured by the Eulerian observer $\mathcal{O}_{0}$,  written as \cite{lichnerowicz1967relativistic}:
\be
E_{\alpha} = \left( 0 , \frac{1}{N} \left[  \frac{\partial A_{t}}{\partial r} + \omega \frac{\partial A_{\phi}}{\partial r}\right ] , \frac{1}{N} \left[  \frac{\partial A_{t}}{\partial \theta} + \omega \frac{\partial A_{\phi}}{\partial \theta}\right ]   , 0 \right),
\ee
and the magnetic field  given by:
\be
\hspace{-2cm} B_{\alpha} = \left( 0 , \frac{1}{B r^{2} \sin \theta} \frac{\partial A_{\phi}}{\partial \theta}, - \frac{1}{B \sin \theta} \frac{\partial A_{\phi}}{\partial r} , 0  \right),
\ee
\\
%with $N^{\phi} (r, \theta)$, $N (r, \theta)$, $\Psi$  metric potentials (for more details see Refs. \cite{Bonazzola:1993zz, Bocquet:1995je, Chatterjee:2014qsa}).  
with $A_{t}$ and $A_{\phi}$ the two non-zero components (for a poloidal magnetic field) of the electromagnetic four-potential $A_{\mu} = (A_{t}, 0 , 0 , A_{\phi})$.

A coordinate-independent characterization of the stellar equator can be done by calculating the circumferential equatorial radius ($\mathsf{\theta = \pi/2}$):
\be 
R_{circ} := \frac{1}{2 \pi} C,
\label{RCIRC}
\ee
with $C$ being the proper length of the circumference of the star in the equatorial plane. So, one obtains:
\be
R_{circ} = \frac{1}{2 \pi} \int_{0}^{2\pi}
\sqrt{\gamma_{\phi\phi}} d\phi =\int_{0}^{2\pi} B(r,\theta)r\sin \theta d\phi  = r_{eq} B(r_{eq},\pi/2),
\label{RCIRCFINAL}
\ee
where $r_{eq}$ is the equatorial coordinate radius.

\section{Spherical vs. deformed solutions}

We start by using the MBF model to calculate EoS' of charge-neutral beta-equilibrated hadronic matter at zero temperature including the effects of Landau quantization. This formalism in then applied to describe (non-rotating) magnetic neutron stars by solving the Einstein-Maxwell equations using the LORENE C++ library.

First, it is important to point out that as the magnetic field limit for stable solutions within this approach is of the order of $\sim 10^{18}\,\mathrm{G}$, the inclusion of strong magnetic fields on the calculations of the EoS of hadronic matter does not play a significant role for the global properties of the stars, as was already discussed in references \cite{Chatterjee:2014qsa,Franzon:2015sya}. This stems from the fact that the effects of strong magnetic fields in the equation of state become very significant only for magnetic fields of about $\sim 5\times10^{18}\,\mathrm{G}$ \cite{Dexheimer:2011pz}, which are higher than the central magnetic fields generated in the poloidal configurations considered here.

\begin{figure}[!ht]  
  \centering
  \includegraphics[width=.9\linewidth]{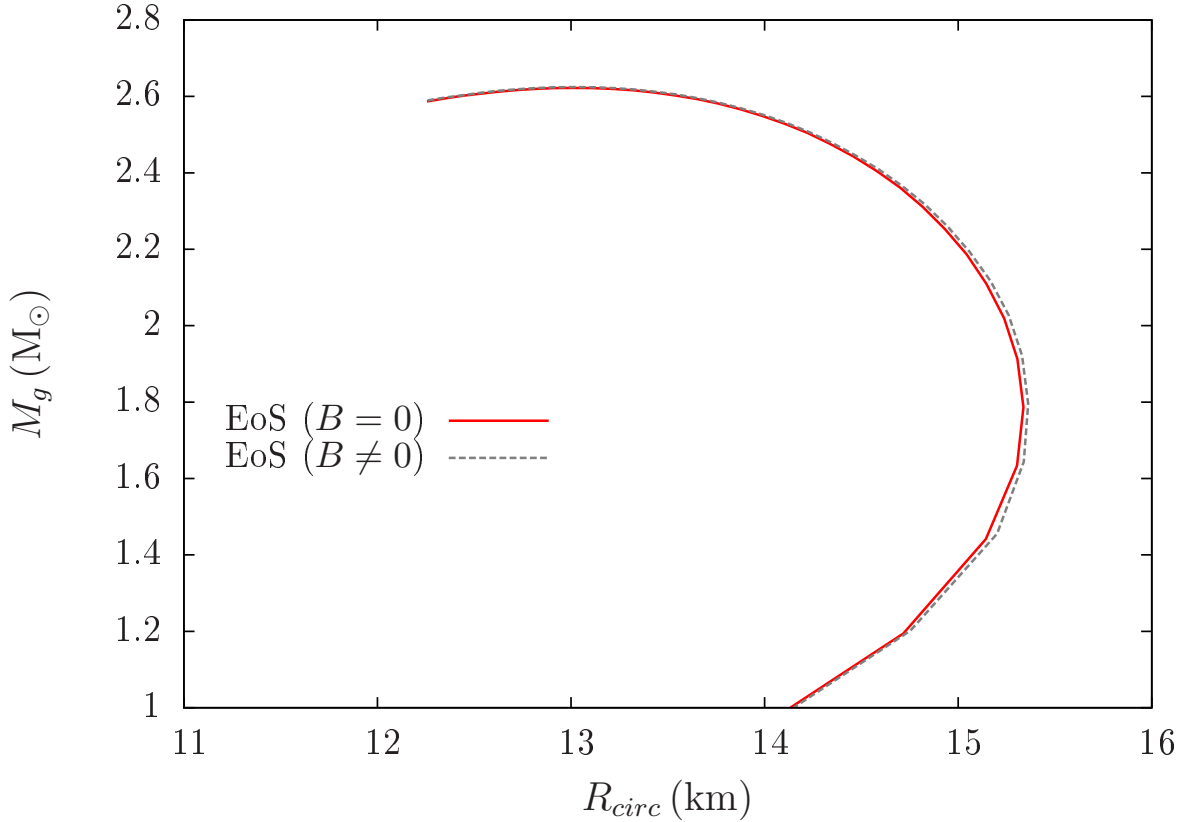}
  \caption{Mass-radius diagram for the parametrization
$\zeta = 0.040$ of the MBF model with a current
function $j_0=3.5\times10^{15}\,\mathrm{A/m^2}$ 
including and not including magnetic effects on the EoS. The vertical axis shows the gravitational mass and the horizonal one the circular radius.}
\label{eosBLORENE}
\end{figure}

That being said, we have calculated the mass-radius diagram using the MBF model with and without magnetic field effects on the EoS of nucleonic matter only and also for matter including hyperons. We confirmed that the global properties of stars are essentially the same in both cases (including hyperons or not). This can be seen in Figure \ref{eosBLORENE}, which shows the mass-radius diagram only for a specific parametrization of the MBF model (with nucleons and leptons) and a specific current function used in the solution of the Einstein-Maxwell equations (2D solutions). Our results are in agreement with the NJL model calculations used to describe quark stars \cite{Chatterjee:2014qsa}. In the case of Ref. \cite{Franzon:2015sya}, where a chiral model was used to describe hybrid stars, this is not exactly the case, because the baryonic anomalous magnetic moments (AMM) enhance magnetic field effects and, therefore, have some influence on the macroscopic properties of stars. Since in this work AMM effects are not included, all subsequent analysis of macroscopic stellar properties is done without taking into account the effects of magnetic fields on the EoS of the MBF model.
However, it is important to stress that, as magnetic field effects modify the population of stars, cooling processes must also be altered \cite{Dexheimer:2011pz,Sinha:2015bva,Tolos:2016hhl}. 

In the past, several authors have described magnetic neutron stars 
including magnetic field effects only on the EoS and computing the macroscopic structure of stars by solving the spherical isotropic Tolman-Oppenheimer-Volkoff (TOV) equations, under the assumption that the deformation of magnetic stars would be small. However, recent self-consistent calculations for magnetic stars  \cite{Franzon:2015sya} have proven this assumption wrong, since the stars can be deformed by more than $50\%$ for a poloidal magnetic field distribution that reproduces strong central magnetic fields up to $\sim 10^{18}\,\mathrm{G}$. 

\begin{figure}[!ht]  
  \centering
  \includegraphics[width=.9\linewidth]{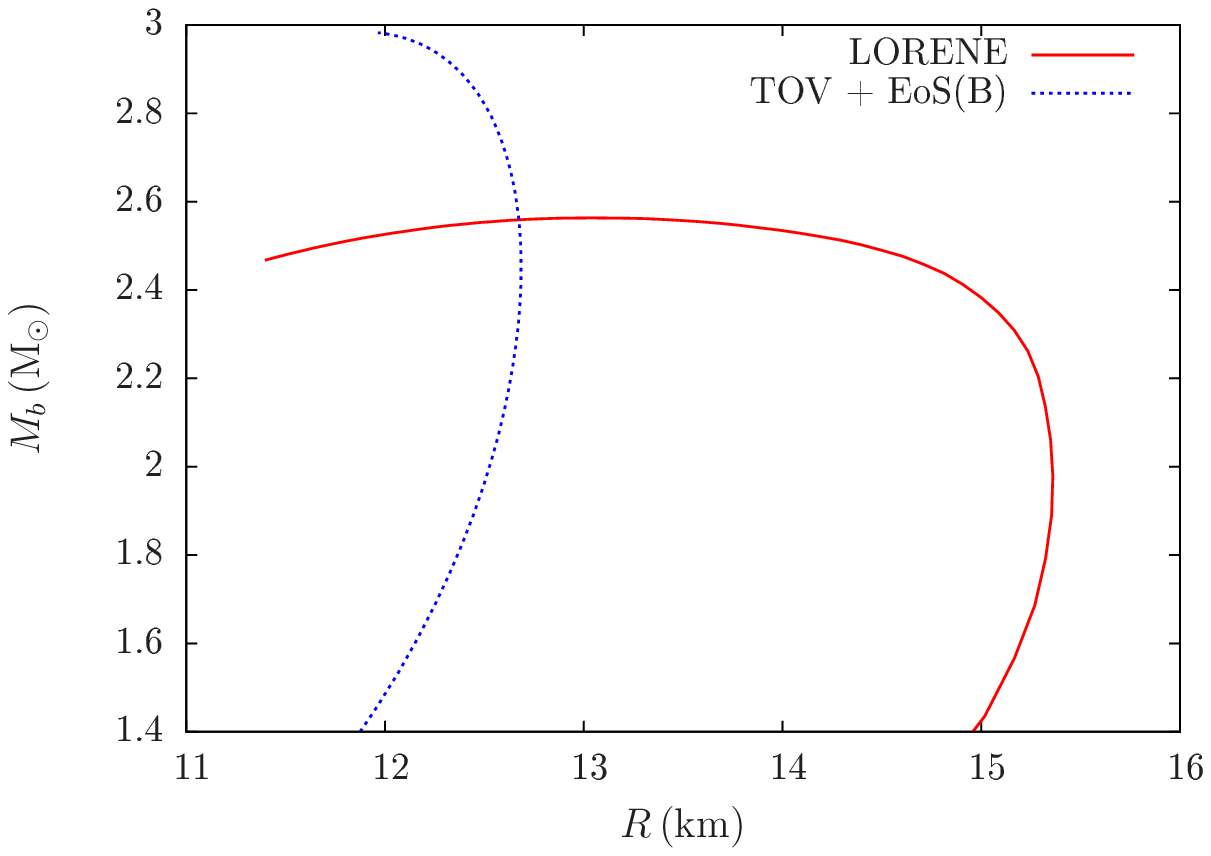}
  \caption{Comparison of the baryon mass-radius relation for magnetic hyperon stars described by the LORENE (red) and TOV (blue) solutions, using the magnetic field configuration $j_0=3.5\times10^{15}\,\mathrm{A/m^2}$ in the first case. The vertical axis describes the baryon mass of the stars, and the horizontal axis the radius (equatorial radius for LORENE). Both calculations were made for the $\zeta=0.040$ parametrization of the MBF model.}
\label{tovMb_wrong}
\end{figure}

To exemplify this, we compare the mass-radius relations for the 2D  solutions described in the previous section and the TOV solutions for the highest magnetic field configuration considered in this work, using an  electric current amplitude  $j_0=3.5\times 10^{15}\mathrm{A/m^2}$ in the first case. 
%First the mass-circular radius relation obtained by solving the Einstein-Maxwell equations for a family of stars with different central densities. 
For the TOV solution, we make use of the magnetic field profile calculated from the 2D solutions 
(given by the maximum mass configuration) in the pure magnetic field contribution, which is added isotropically to the EoS. Note that this is not correct, since the pure magnetic field contribution enters with different signs in different directions in the energy-momentum tensor; however, this is a frequently used assumption in the literature. %The modified EoS in then used in the TOV equations for comparison and quantification of the associated error on  using TOV equations for highly magnetized stars.
%the EoS with magnetic field effects.  The TOV equations are calculated using the pressure component that is perpendicular to the magnetic field and by comparing the two formalism for the baryon and gravitational mass-radius diagrams, from Figures \ref{tovMb_wrong} and \ref{tovMg_wrong} respectively, one can quantify the error associated to the use of the latter formalism to describe such objects. 

From Figure \ref{tovMb_wrong}, one can check that not taking magnetic field effects into account for the macroscopic structure of stars leads to an overestimation of the maximum mass allowed by a specific EoS, as well as to an underestimation of the equatorial radius of stars. In particular, for the parametrization $\zeta=0.040$ of the MBF model with hyperons, we obtain maximum baryon masses of $M_{b}^{max}=2.57\,\mathrm{M_{\odot}}$ and $M_{b}^{max}=2.98\,\mathrm{M_{\odot}}$ for 2D and TOV solutions with magnetic fields, respectively. 
A similar comparison can be done for the radii of a  $M_{b}=1.4\,\mathrm{M_{\odot}}$ star, from which we calculate  $R_{1.4\mathrm{M_{\odot}}}=14.95\,\mathrm{km}$ and $R_{1.4\mathrm{M_{\odot}}}=11.88\,\mathrm{km}$, again for deformed and TOV solutions with magnetic fields, respectively.  
Here it is important to stress that the radial comparison is done between the equatorial radius for the 2D solution and the 
isotropic radius of the magnetic TOV result.

\begin{figure}[!ht]  
  \centering
  \includegraphics[width=.9\linewidth]{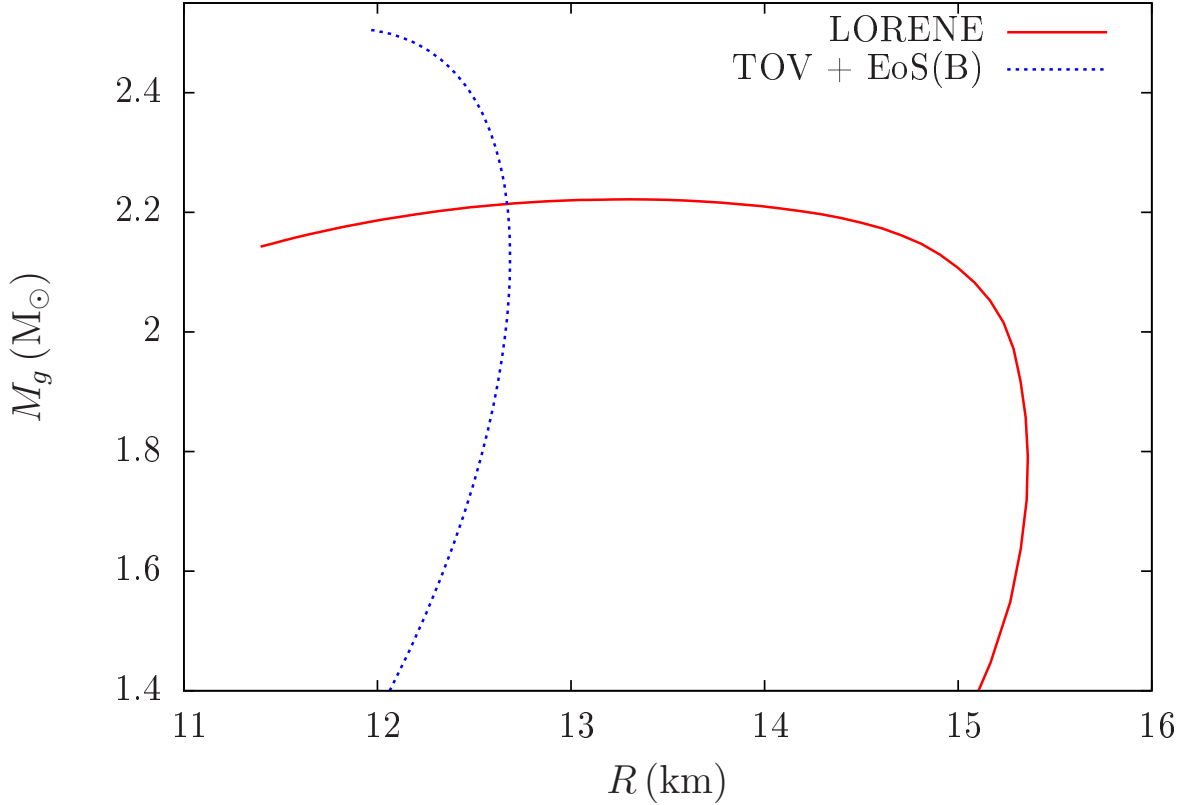}
  \caption{Same as Figure \ref{tovMb_wrong}, but showing the gravitational mass of stars on the vertical axis.
}
\label{tovMg_wrong}
\end{figure}

An analogous analysis can be performed for the gravitational mass, as it is shown in Figure \ref{tovMg_wrong}. For the same  parametrization of the nuclear model, maximum gravitational masses of $M_{g}^{max}=2.23\,\mathrm{M_{\odot}}$ and $M_{g}^{max}=2.50\,\mathrm{M_{\odot}}$ are estimated for 2D and TOV solutions with magnetic fields, respectively. 
The radius of the $M_{g}=1.4\,\mathrm{M_{\odot}}$ stars are  $R_{1.4\mathrm{M_{\odot}}}=15.11\,\mathrm{km}$ and $R_{1.4\mathrm{M_{\odot}}}=12.05\,\mathrm{km}$, again, for the 2D and magnetic TOV cases, respectively.  

We now focus on the maximum baryon mass star obtained from the 2D LORENE solution, with a maximum baryon mass  $M_{b}^{max}=2.57\,\mathrm{M_{\odot}}$. One can check that the corresponding gravitational mass for the magnetic TOV solution is $M_{g}=2.21\,\mathrm{M_{\odot}}$, which is a much smaller value than the maximum solution $M_{g}^{max}=2.50\,\mathrm{M_{\odot}}$ obtained from the magnetic TOV solution, but similar to the correct value obtained for the full LORENE solution $M^{max}_{g}=2.23\,\mathrm{M_{\odot}}$. 
Similarly, we can take the $M_{b}=2.2\,\mathrm{M_{\odot}}$ star and check the  circular radius estimation both with the magnetic TOV solution and the 2D one, obtaining  $R_{2.2\mathrm{M_{\odot}}}=12.64\,\mathrm{km}$ and 
$R_{2.2\mathrm{M_{\odot}}}=15.29\,\mathrm{km}$, respectively. 
From these results, we can conclude that there is an overestimation of the maximum gravitational mass (in the magnetic TOV case) when comparing both maximum solutions, but only a small change for a fixed baryon mass. However, for the equatorial radius of stars, the difference is much more pronounced (even for the same baryon mass), with a difference of $2.65\,\mathrm{km}$.

More generally, the results show that there is an overestimation of $15.95\%$ for the maximum baryon mass and of $12.11\%$ for the gravitational mass by the magnetic TOV approach. The equatorial radius is underestimated by $20.54\%$ for the $1.4\,\mathrm{M_{\odot}}$ baryon mass star and in $20.25\%$ for the $1.4\,\mathrm{M_{\odot}}$ gravitational mass star. Part of the errors in the estimate of global properties of magnetic neutron stars comes from the inappropriate metric used to solve the TOV equations. A spherical metric with equal-sign pressure contributions incorrectly describes the  magnetic field in all directions, allowing for stable more massive stars. When one follows an axi-symmetric approach, part of the magnetic pressure components generate the stellar deformation, and only part of the field effects lead to an increase in the mass of stars.

\begin{figure}%[!ht]
\centering
   \includegraphics[width=1.03\linewidth]{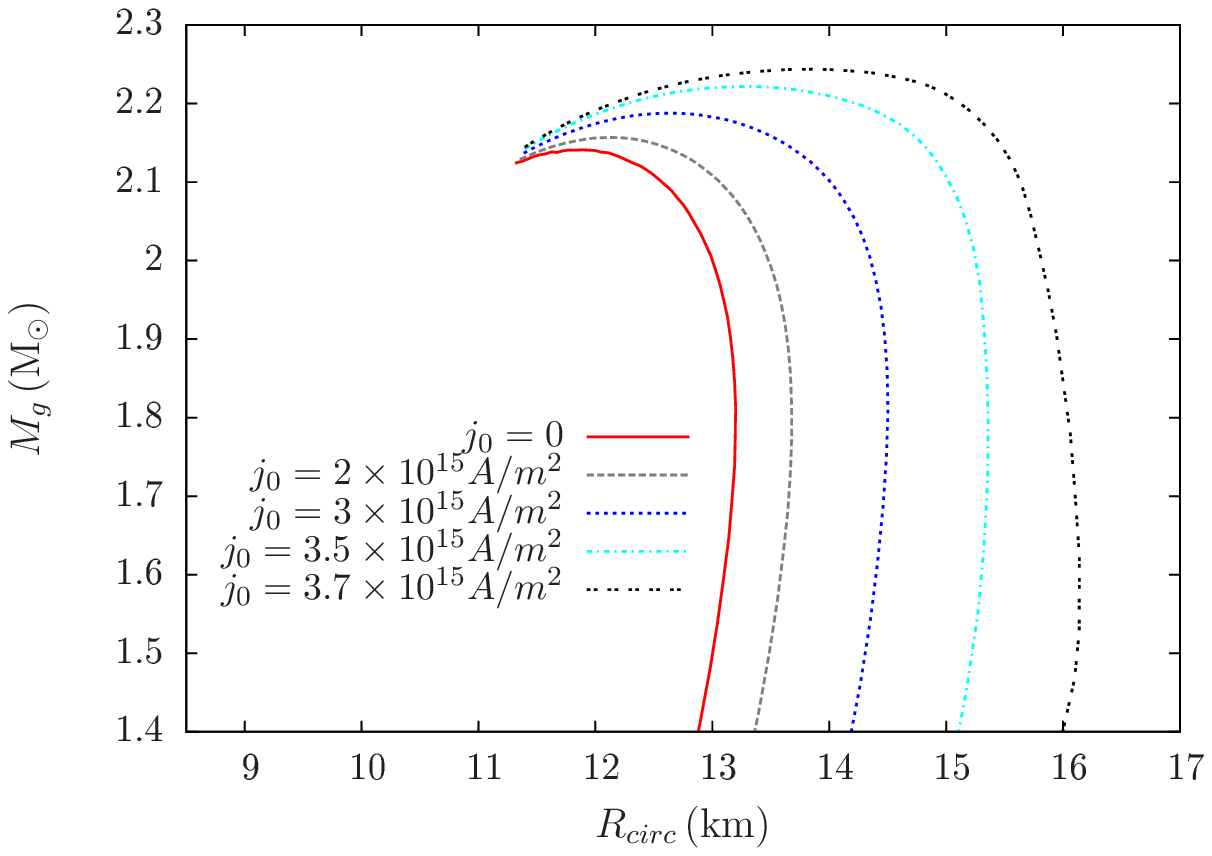}
  %\captionof{figure}{$\mathbf{(g_{\varrho n}/m_{\varrho})^2}$}
    \caption{Same as Figure \ref{tovMg_wrong}, but using different electric current amplitudes $j_0$.
    %Mass-radius relation of magnetic stars. The vertical axis indicates the mass of the stars and the horizontal axis, the radius.  For the case in which the Einstein-Maxwell equations are solved, the radius corresponds to the circular one. The panel to the left shows the diagram for different initial currents. The panel to the right compares the mass-radius relation for the case in which the magnetic effecs are only taken into account in the EoS (using the TOV equations) and by solving the Einstein-Maxwell equations, that considers the magnetic effects on the structure of the stars.
}\label{tovM}
 \end{figure}

Figure \ref{tovM} shows the impact of different magnetic field configurations on the mass-radius relation, for the parametrization $\zeta= 0.040$ of the MBF model, with hyperons. 
The red full curve is the non-magnetic TOV solution, which gives a maximum gravitational mass $M_g = 2.15\,\mathrm{M_{\odot}}$. The other curves correspond to different choices for the electric current amplitude $j_0$ in the LORENE code, which generate a different magnetic field configuration throughout the star. In particular, the black curve for $j_0=3.7\times10^{15}\,\mathrm{A/m^2}$ corresponds to the maximum central magnetic field configuration $B_c=1.1\times10^{18}\,\mathrm{G}$ (doubled dotted). The results show that magnetic stars present a gravitational mass increase, from $2.15\,\mathrm{M_{\odot}}$ to $2.22\,\mathrm{M_{\odot}}$ for the maximum mass configuration, and a increase of respective radii from $R_{1.4\mathrm{M_{\odot}}}=12.88\,\mathrm{km}$ 
to $R_{1.4\mathrm{M_{\odot}}}=15.98\,\mathrm{km}$, from the non-magnetic to the highest magnetic field configuration.
This is due to the fact that the Lorentz force acts against gravity, allowing stars to support more mass. Also, the increase in the equatorial radius is associated with a more pronounced deformation of stars into oblate form. This is the shape favored by a poloidal magnetic field  distribution, which is assumed in this work. 

%In what follows, we apply different parametrizations of the MBF model in order to 
%investigate stellar solutions with fixed baryon mass $M_b=2.2\,\mathrm{M_{\odot}}$. 
%For the reasons mentioned above, we do not take magnetic field effects on the equation of state into account.

\section{Properties of a (fixed baryon mass) Magnetic Star}

In this section, we investigate the impact of different degrees of stiffness of the EoS on the properties of magnetic neutron stars.
To do so, we fix the baryon mass of stars to $M_b=2.2\,\mathrm{M_{\odot}}$ and calculate their properties for different values of the $\zeta$ parameter of the MBF model.%, as well as different values of the electric current amplitude.

As shown in Section III (see Eq. (\ref{currentEOS})), the electric current which gives rise to a magnetic field distribution which depends both on the equation of state and the electric current amplitude. This means that, for a fixed electric current amplitude, the magnetic dipole moment $\mu$ of stars, as well as their surface and central magnetic field strengths vary for different parametrizations. The same is true if we fix the parametrization of the model and vary the electric current amplitude. 
The latter topic has been extensively explored in previous works about magnetic neutron stars \cite{Bonazzola:1993zz,Bocquet:1995je,Cardall:2000bs,Chatterjee:2014qsa,Franzon:2015sya,Franzon:2016iai}. For this reason, in this section we focus on the effects of different parametrizations of the MBF model, although a short discussion regarding different electric current amplitudes is also presented.

Table \ref{stars_models} shows the magnetic dipole moment, surface and central magnetic fields, and the central density for a $M_b=2.2\,\mathrm{M_{\odot}}$ star, for different choices of the many-body forces parameter $\zeta$ and electric current amplitude $j_0$. We vary  the many-body forces parameter of the MBF model in order to cover the whole accepted experimental range of nuclear matter properties at saturation, $\zeta=0.040-0.129$; and the electric current amplitude is varied in order to reach the maximum possible magnetic configuration in at least one of the parametrizations, $j_0=(1.0-3.5)\times10^{15}\,\mathrm{A/m^2}$. The results shown in this table are for nucleonic stars, although they are roughly the same as the ones for hyperon stars, as is going to be discussed later, when analyzing the population inside the stars.

\begin{table*}[t]
  \caption{\label{stars_models} Characteristics of a magnetized nucleonic star with $M_b=2.2\,\mathrm{M_{\odot}}$ for different 
  many-body forces parameter values and magnetic field distributions. 
The columns are, respectively, the many-body forces parameter $\zeta$, the electric current amplitude $j_0$, the dipole moment of the stars $\mu$, the surface and central magnetic fields, and the central density for each choice of parameters.}
\begin{center}
\begin{tabular}{cccccc}
 \hline
$\zeta$ & $j_0\,(10^{15}\mathrm{A/m^2})$ &  $\mu\,(10^{32}\mathrm{Am^2})$ & $ B_s\,(10^{17}\mathrm{G}) $ & $ B_c\,(10^{17}\mathrm{G}) $ & $\rho_c \,(\mathrm{fm^{-3}})$   \\
  \hline \hline
    %&  &  &  &  \\
 0.040 & $0$ & n.a.  & n.a.  & n.a. & 0.376  \tabularnewline   
 0.040 & $1.0$ & $0.57$  & $1.03$  & $3.48$ & 0.375  \tabularnewline
 0.040 & $2.0$ & $1.21$  & $1.47$  & $4.66$ & 0.364  \tabularnewline
 0.040 & $3.0$ & $2.09$  & $2.70$  & $7.03$ & 0.335  \tabularnewline
 0.040 & $3.5$ & $2.88$  & $3.80$  & $8.09$ & 0.299  \tabularnewline

\hline

 0.059 & $0$ & n.a.  & n.a.  & n.a. & 0.432  \tabularnewline   
 0.059 & $1.0$ & $0.52$  & $0.67$  & $2.44$ & 0.432 \tabularnewline
 0.059 & $2.0$ & $1.09$  & $1.47$  & $4.91$ & 0.420 \tabularnewline
 0.059 & $3.0$ & $1.86$  & $2.64$  & $7.39$ & 0.389 \tabularnewline
 0.059 & $3.5$ & $2.47$  & $3.60$  & $8.52$ & 0.354 \tabularnewline

\hline

 0.085 & $0$ & n.a.  & n.a.  & n.a. & 0.519  \tabularnewline    
 0.085 & $1.0$ & $0.46$  & $0.68$  & $2.62$ & 0.517 \tabularnewline
 0.085 & $2.0$ & $0.96$  & $1.48$  & $5.26$ & 0.503 \tabularnewline
 0.085 & $3.0$ & $1.60$  & $2.59$  & $7.88$ & 0.469 \tabularnewline
 0.085 & $3.5$ & $2.07$  & $3.42$  & $9.11$ & 0.434 \tabularnewline

\hline

 0.129 & $0$ & n.a.  & n.a.  & n.a. & 0.676  \tabularnewline    
 0.129 & $1.0$ & $0.38$  & $0.70$  & $2.92$ & 0.668 \tabularnewline
 0.129 & $2.0$ & $0.80$  & $1.50$  & $5.84$ & 0.651 \tabularnewline
 0.129 & $3.0$ & $1.30$  & $2.55$  & $8.72$ & 0.610 \tabularnewline
 0.129 & $3.5$ & $1.63$  & $3.27$  & $10.1$ & 0.573 \tabularnewline

  \hline\hline
  
  \end{tabular}
\end{center}
\end{table*}

From Table \ref{stars_models}, one can see that higher values of the electric current amplitude lead to higher magnetic moment for stars, as well as more intense magnetic field distributions (surface $B_s$ and central $B_c$). These results come essentially from the fact that a higher surface current can generate stronger magnetic fields and, consequently, more magnetized stars. 
On the other hand, the magnetic fields decrease the central density of stars, similarly to the centrifugal force in rotating stars. As is going to be discussed in the next section, this effect has a dramatic impact on the particle populations of magnetic stars. 

\begin{figure}%[!ht]
\centering
   \includegraphics[width=1.03\linewidth]{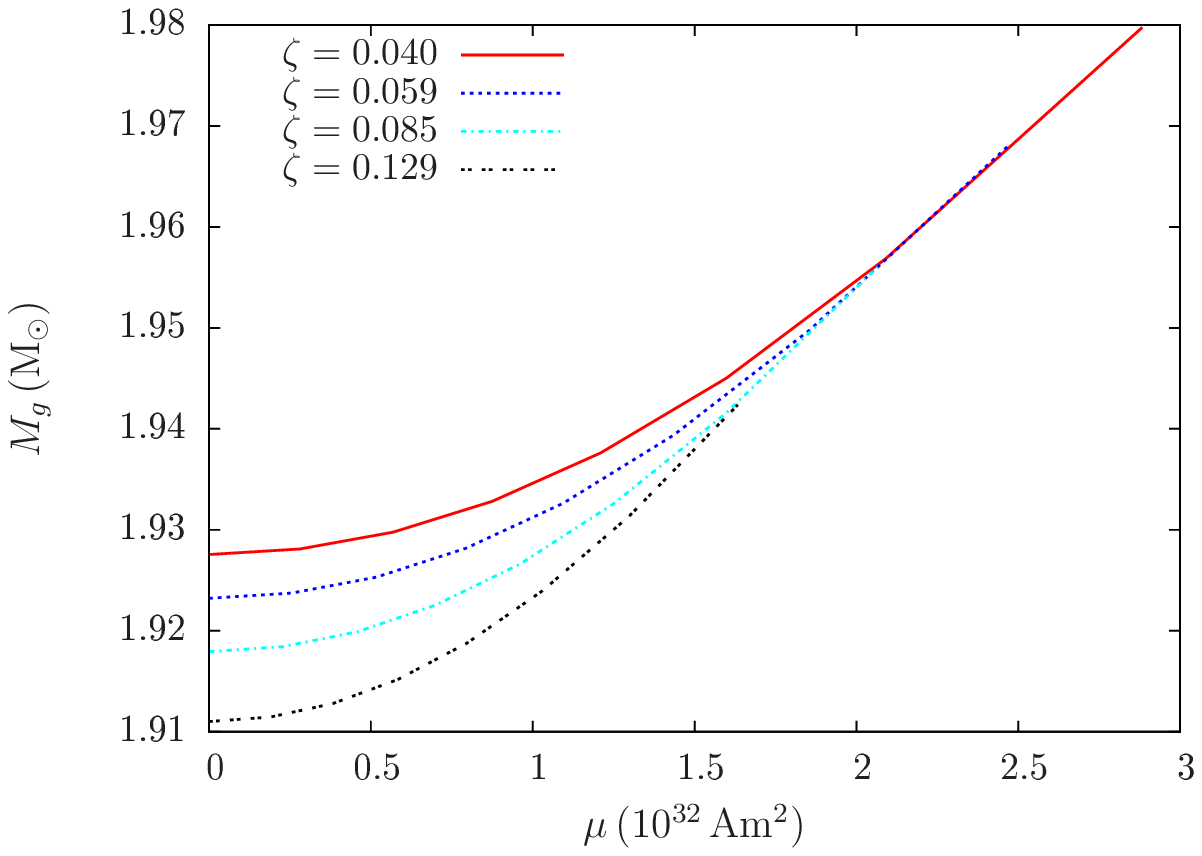}
    \caption{Gravitational mass dependence on the many-body forces contributions and  different magnetic field configurations. 
The vertical axis is the gravitational mass and the horizontal axis is the magnetic moment of the corresponding magnetic field configuration for a $M_b = 2.2\,\mathrm{M_{\odot}}$ star.}
\label{mg}
 \end{figure}

\begin{figure}[!ht]  %colocar o gráfico aqui ou no topo da página
  \centering
  \includegraphics[width=1.05\linewidth]{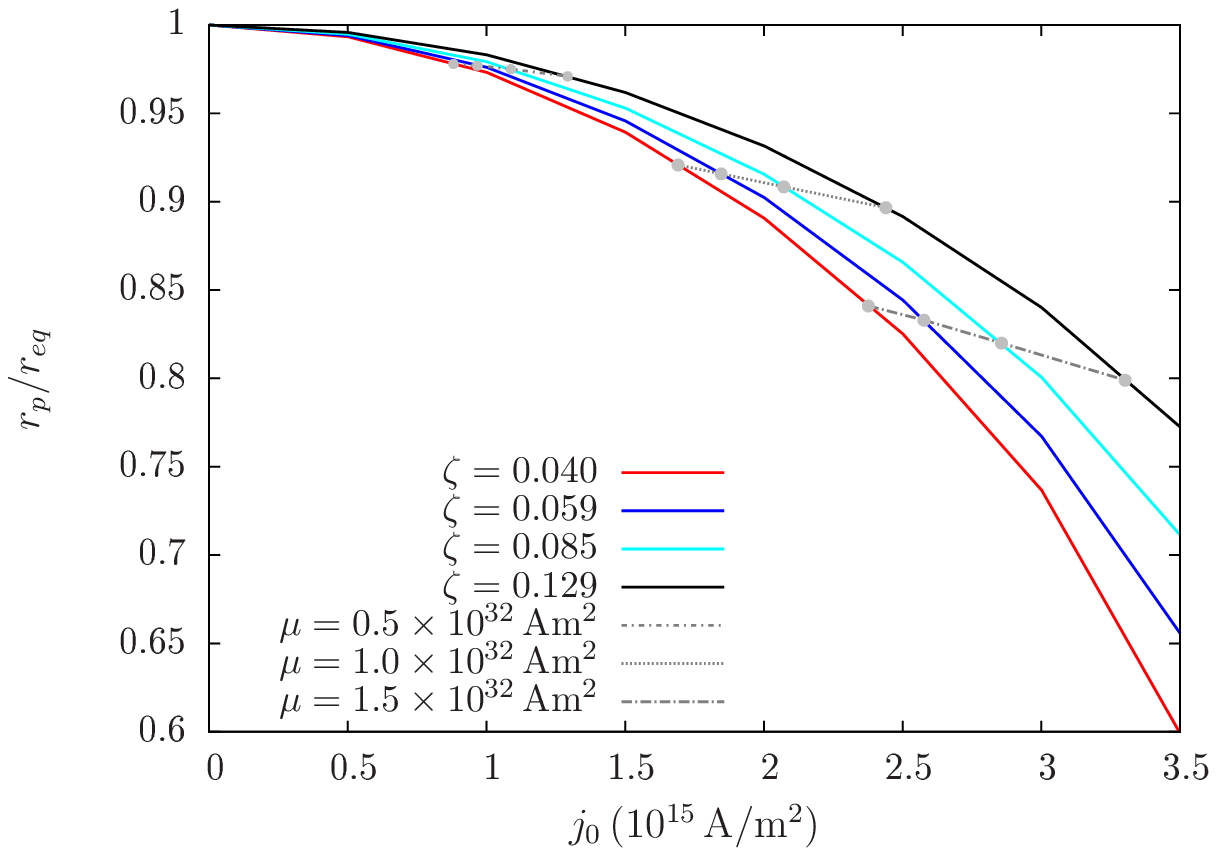}
    \caption{Stellar deformation as function of the electric current amplitude for equations of state with different many-body forces contributions. 
The vertical axis is the ratio between the polar and equatorial coordinate radii and the horizontal axis is the electric current amplitude for 
$M_b = 2.2\,\mathrm{M_{\odot}}$ star. The gray lines connecting the dots correspond to results with fixed magnetic moment.}
\label{reqrp}
\end{figure}

Figures \ref{mg} and \ref{reqrp} show, respectively, the dependence of the gravitational mass and deformation of stars on the many-body forces parameter, respectively.  Figure \ref{reqrp} also shows different magnetic field configurations, associated with the magnetic dipole moment of the stars $\mu$. As already discussed, the higher the magnetic dipole moment and  electric current amplitudes the higher the Lorentz force, which acts against gravity, enhancing the gravitational mass, as well as the equatorial radius of stars, which become more oblate (a $rp/r_{eq}$ ratio farther from $1$).

  % rp/req

From the equation of state point of view, smaller values of the many-body forces parameter have a shielding effect on the scalar couplings, maximizing the vector meson contributions and, thus, generating stiffer EoS's (see Ref. \cite{Gomes:2014aka} for more details). Consequently, they allow for more massive and larger stars (see Figure \ref{mg}). 
Therefore, for a fixed magnetic field distribution, small values of $\zeta$ generate higher gravitational masses.
%, as the repulsive matter interaction is stronger for itself, besides the magnetic contributions. 
Nevertheless, for larger values of the magnetic dipole moment, the magnetic field contribution dominates and the EoS effects cannot be seen.
For the highest magnetic field configuration reported in this work, the associated gravitational mass for the different parametrizations ranges from $1.98\,\mathrm{M_{\odot}}$ (for $\zeta=0.040$) to $1.94\,\mathrm{M_{\odot}}$ (for $\zeta=0.129$) for $j_0=(1.0-3.5)\times10^{15}\,\mathrm{A/m^2}$ keeping $M_b=2.2\,\mathrm{M_{\odot}}$.

\begin{figure}[!ht]  %colocar o gráfico aqui ou no topo da página
  \centering
  \includegraphics[width=1.1\linewidth]{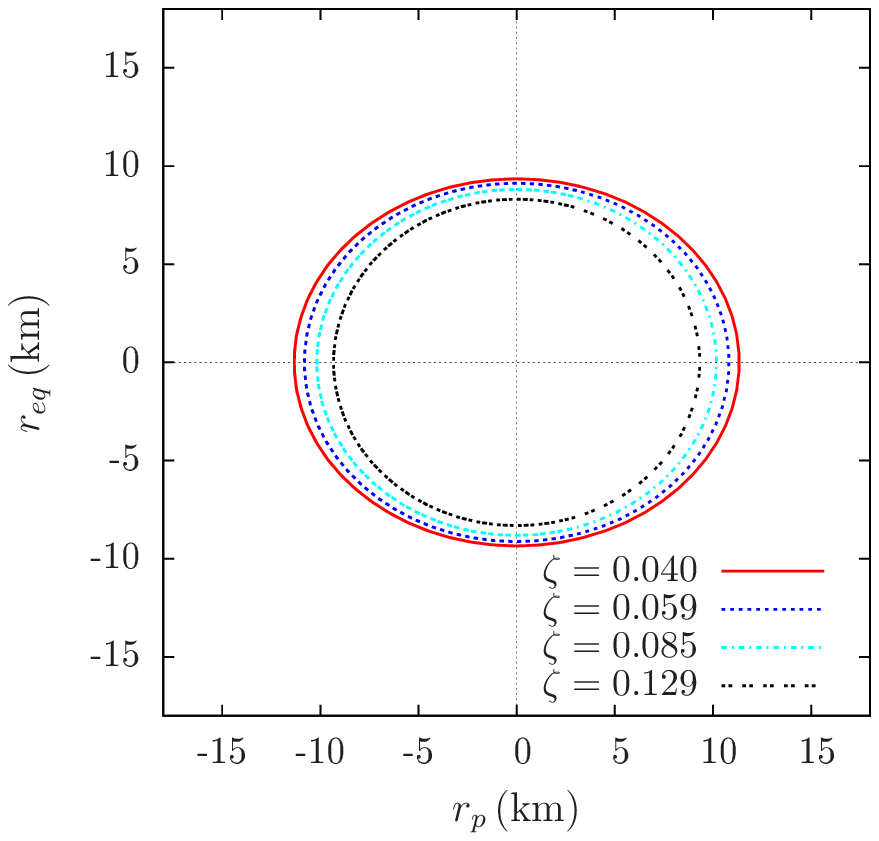}
  \caption{Two-dimensional cut  of a $M_b = 2.2\,\mathrm{M_{\odot}}$ star , for the highest magnetic field configuration $j_0=3.5\times10^{15}\,\mathrm{A/m^2}$ showing the dependence on the many-body forces parameter. The vertical and horizontal axes are, respectively, the equatorial and polar coordinate radii.}
\label{deformation_z}
\end{figure}

As one can see in Figure \ref{deformation_z}, the smaller the $\zeta$, the larger the radius of stars. Note that, although the mass in this case is larger, the stellar volume scales as $R^3$. As a consequence, the central density is lower for smaller values of the many-body forces parameter (as shown  in Table \ref{stars_models}). 
In this way, we can identify relations between radius, central density, and central magnetic field with the deformation of stars. 
In particular, the most deformed configuration (for $\zeta=0.040$ and  
$j_0=3.5\,\times10^{15}\mathrm{A/m^2}$) generates a radius ratio  $r_p/r_e=0.6$,  as one can see in Figure \ref{reqrp}.

This is not a straightforward result since the 2D calculation of all  star properties depends both on the EoS and on the magnetic field configuration, given by the electric current amplitude. A stiffer EoS allows for higher radii and, consequently, lower central densities and central magnetic fields. However, these stars are  less compact and, consequently, more easily deformed. In principle, one could expect the softer EoS to be the one that generates more deformed stars due to the higher central magnetic field. Nevertheless, as softer EoS stars have smaller radii, they are more compact and, hence, more difficult to deform.

% deformacao zeta

Figure \ref{reqrp} also shows the isocontours for fixed values of the magnetic moment of stars as a function of the many-body parameter and the electric current amplitude. The results show that it is necessary to increase the electric current amplitude in order to reproduce the same dipole moment for a fixed baryon mass star, if this star is described by a soft EoS. Here, again, the determination of the magnetic moment depends both on the EoS (which is determinant for the radius) and on the magnetic field distribution, which comes from $j_0$. 
In order to have the same magnetic moment, the softer EoS must compensate its smaller radius by increasing the magnetic field, generating slightly more deformed stars for the case of fixed magnetic moment.% (CHECK STATEMENT: SO IN THE CASE OF SAME J0, THE RADIUS IS THE DETERMINANT PART OF THE DEFORMATION, AND FOR FIXED MU IT IS THE B?).

%Finally, for the same reason as the stars are less deformed due to the lower central densities for stiff parametrizations of the EoS, the magnetic moment of the stars are smaller for small values of $\zeta$, as one can see in Table \ref{Hmodels}.

%% perfil densidade
\begin{figure}[!ht]  %colocar o gráfico aqui ou no topo da págin]
  \centering
  \includegraphics[width=.9\linewidth]{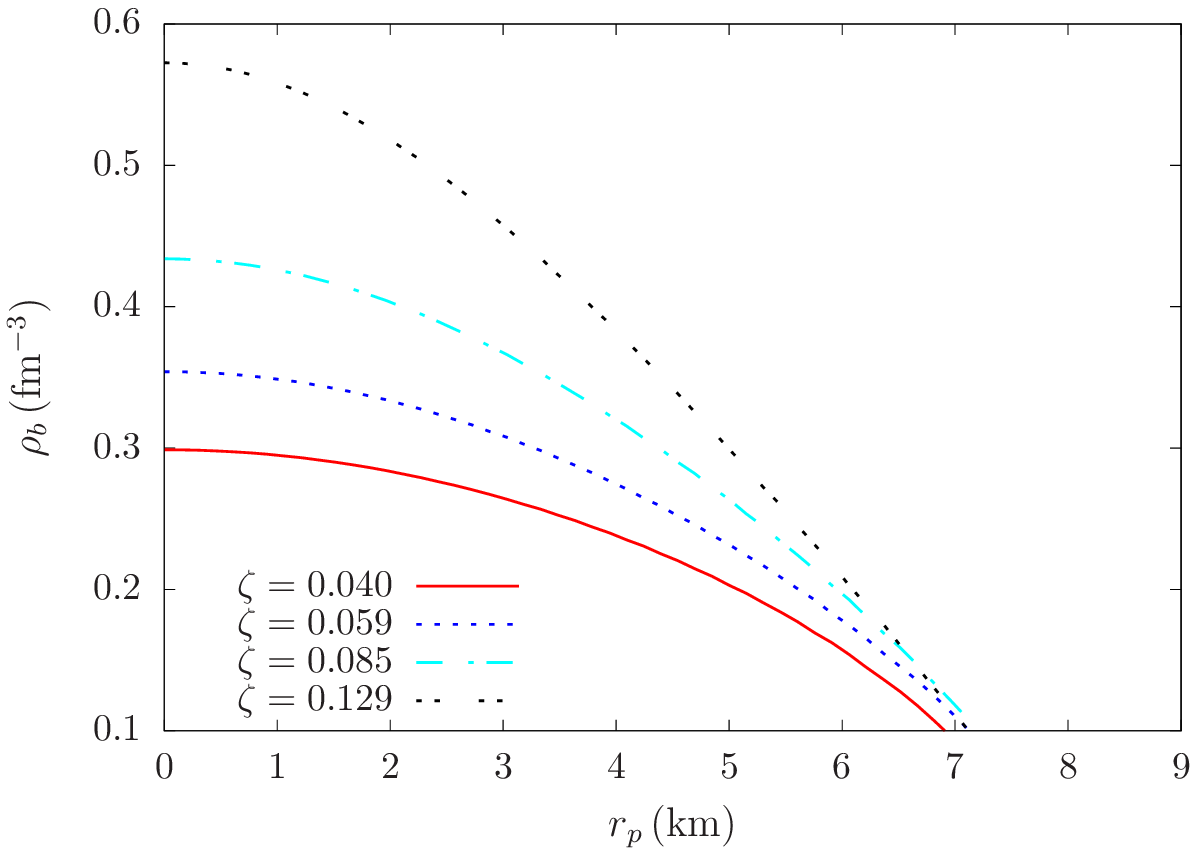}
  \caption{Density profile for $M_b=2.2\,\mathrm{M_{\odot}}$ stars with  $j_0=3.5\times10^{15}\,\mathrm{A/m^2},$ for different choices of the many-body forces parameter $\zeta$. The vertical and horizontal axes are, respectively, the baryon density and the polar radius of stars.}
\label{rho_perfil}
\end{figure}

In the following, we focus our attention on the interior of a star with $M_b=2.2\,\mathrm{M_{\odot}}$,
for the highest magnetic field configuration $j_0=3.5\times10^{15}\,\mathrm{A/m^2}$ in order to check the impact of different many-body forces contributions on their magnetic field and densities profiles.
The baryon density distribution as a function of the polar radius is shown in Figure \ref{rho_perfil}, for different choices of the $\zeta$ parameter. 
As already mentioned, the stiffest EoS generates larger radii and  lower central densities. Because of the larger stellar radii produced by the stiffer EoS, it is possible to see a crossing for the density curves. A similar but enhanced result is found for the density profile in equatorial direction, although it is important to stress that the density distribution is anisotropic for magnetic stars. The poloidal magnetic field distribution makes stars oblate, i.e., more 
flattened in the polar direction and expanding in the equatorial direction \cite{Franzon:2015sya}.

Another important point to make regarding the stellar baryon density distribution is that the Lorenz force reverses its direction along the equatorial plane of magnetized stars at some distance from the center, but still inside the star, as already pointed out in Refs. \cite{Cardall:2000bs,Franzon:2015gda}. Note that, depending on the magnetic field distribution (which in turn depends on the EoS), this might even lead to an off-center maximum baryon density \cite{Franzon:2015sya,Franzon:2016iai}.

\begin{figure}[!ht]  %colocar o gráfico aqui ou no topo da págin]
  \centering
  \includegraphics[width=.9\linewidth]{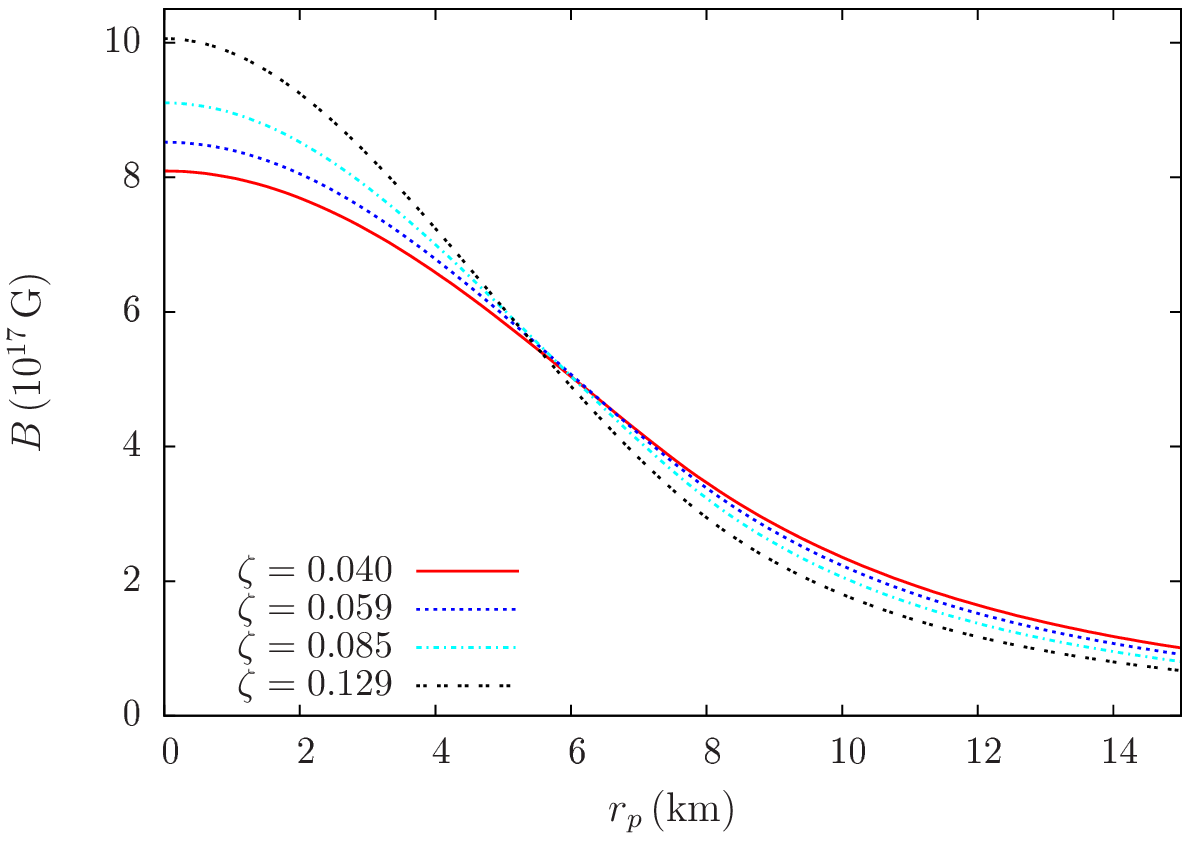}
  \caption{Magnetic field profile for $M_b=2.2\,\mathrm{M_{\odot}}$ stars with  $j_0=3.5\times10^{15}\,\mathrm{A/m^2}$, for different choices of the many-body forces parameter $\zeta$. The vertical and horizontal axes are, respectively, the magnetic field and the polar radius of stars. }
\label{B_profile}
\end{figure}

%Analogously, one can also check the magnetic field profile as function of the polar radius for the same stars. 
As already mentioned, the intensity of the central magnetic field is directly related to the central density reached inside the stars and, hence, higher $\zeta$ values allow for more (centrally)  magnetized stars (see Figure \ref{B_profile}). In particular, for a fixed electric current amplitude configuration, the central magnetic field varies from $8\times10^{17}\,\mathrm{G}$ for the $\zeta=0.040$ parametrization, to $1.01\times10^{18}\,\mathrm{G}$ for the $\zeta=0.129$ case. For a detailed discussion on poloidal magnetic field profiles in magnetic stars, see Ref. \cite{Dexheimer:2016yqu}.

\section{Particle Population}

\begin{figure}[!ht]  %colocar o gráfico aqui ou no topo da página
  \centering
  \includegraphics[width=1.\linewidth]{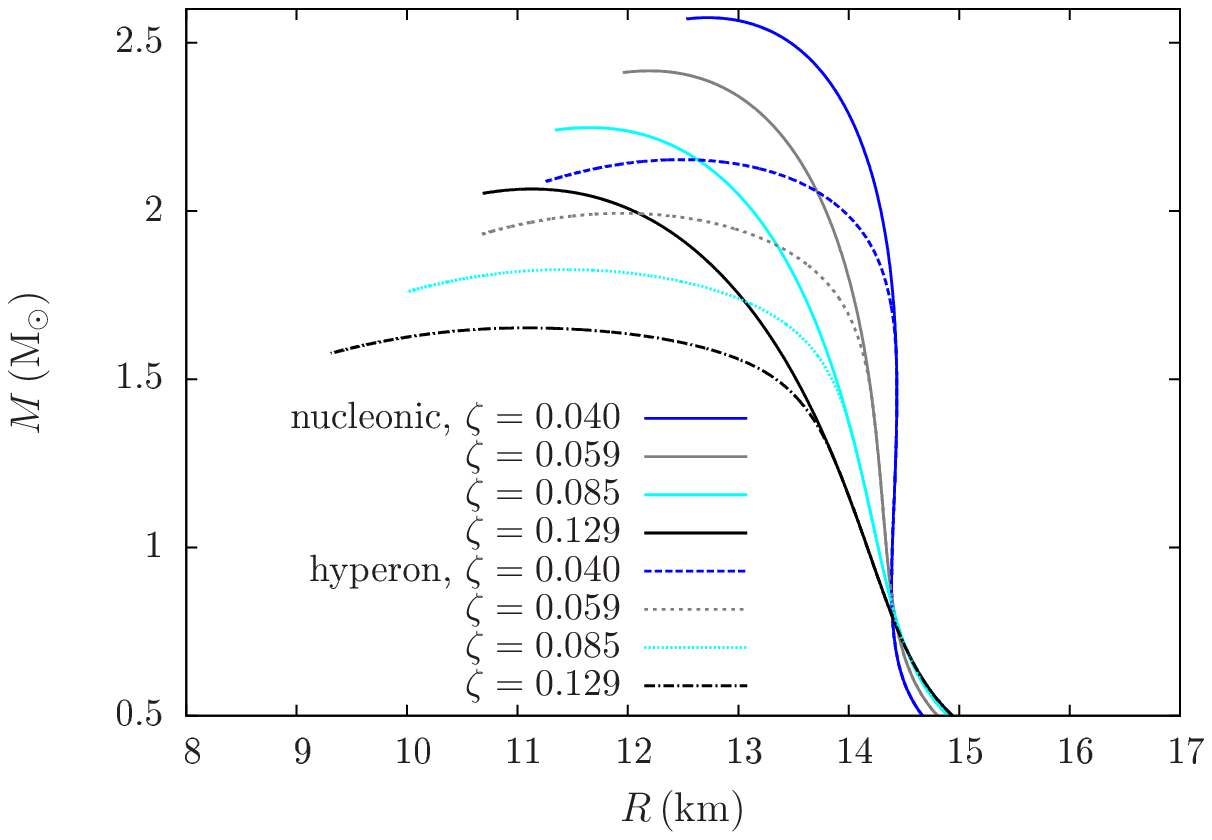}
  \caption{Mass-radius relation for non-magnetic stars for all the parametrizations of the MBF model used in this work. The vertical axis is the gravitational mass and the horizontal axis is radius of stars. Nucleonic stars are displayed with full lines and the hyperon stars, with dashed lines.}
\label{tov_B0}
\end{figure}

As already discussed in the introduction, the global properties of non-magnetic stars are strongly dependent on the particle population and, in particular, on the potential appearance of exotic degrees of freedom at high densities.
In Figure \ref{tov_B0}, we show the mass-radius diagram for the four parametrizations of the MBF model used in this work, for both nucleonic (full lines) and stars that also contain hyperons (dashed lines) not including any magnetic field effects. As one can see, all the parametrizations for nucleonic stars are in agreement with the observational data, but only the first two parametrizations ($\zeta=0.040,\,0.059$) are able to reproduce hyperon stars with masses above $1.97\,\mathrm{M_{\odot}}$. This is the current lower bound including the error in the measurement of massive neutron stars observations \cite{Antoniadis2013}. 
In particular, for nucleonic and hyperon stars, respectively, the maximum masses estimated for each parametrizations (for $B=0$) are $2.57\,\mathrm{M_{\odot}}$ and $2.15\,\mathrm{M_{\odot}}$ for $\zeta=0.040$, 
$2.42\,\mathrm{M_{\odot}}$ and $1.99\,\mathrm{M_{\odot}}$ for $\zeta=0.059$, 
$2.25\,\mathrm{M_{\odot}}$ and $1.83\,\mathrm{M_{\odot}}$ for $\zeta=0.085$ and 
$2.07\,\mathrm{M_{\odot}}$ and $1.65\,\mathrm{M_{\odot}}$ for $\zeta=0.129$ \cite{Gomes:2014aka}, indicating a mass decrease of $\sim 0.42\,\mathrm{M_{\odot}}$ due to the appearance of hyperons.

% tov B= 0 

\begin{figure}[!ht]  %colocar o gráfico aqui ou no topo da página
  \centering
  \includegraphics[width=1.\linewidth]{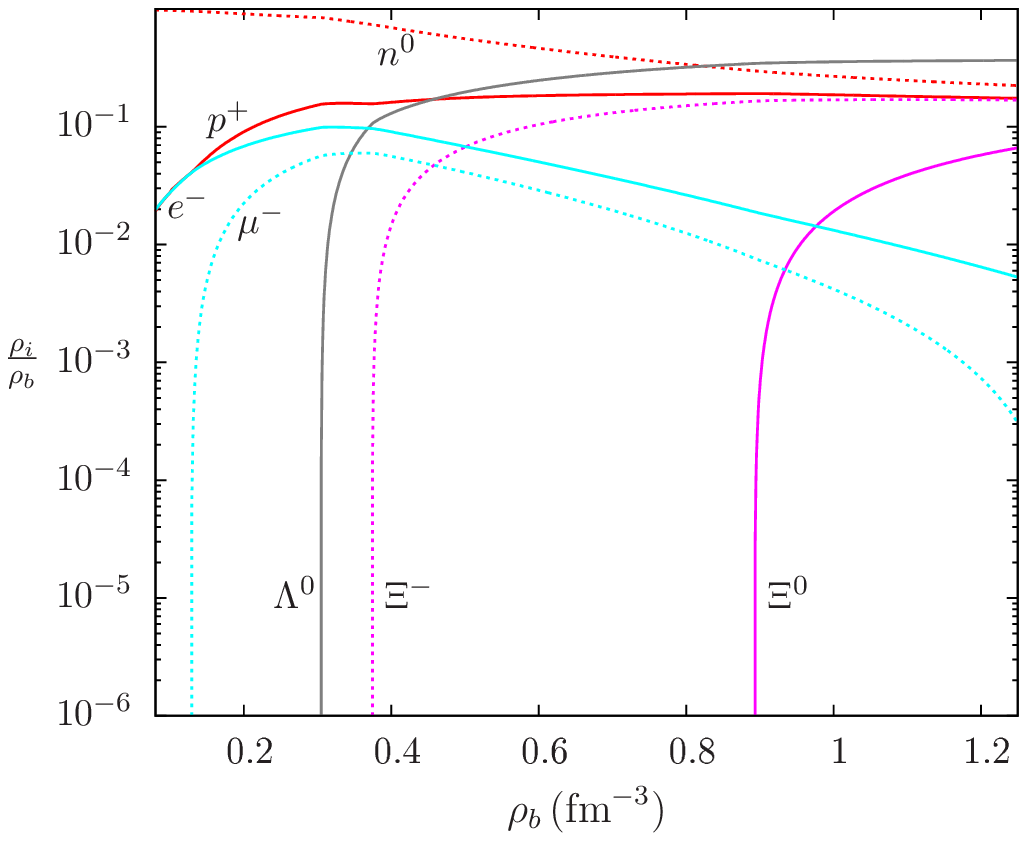}
  \caption{Particles population for the parametrization  $\zeta=0.040$ of the MBF model, for non-magnetic stars.
The vertical axis is the particle fraction normalized by the baryon density, and the horizontal axis is the baryon density.}
\label{pop_l0040}
\end{figure}

Figure \ref{pop_l0040} shows the particle population as function of baryon density for the parametrization $\zeta=0.040$ (the stiffest one), which is the one where hyperons appear at lower densities. 
For this parametrization, the density threshold for $\Lambda$ hyperon appearance is at $0.305\,\mathrm{fm^{-3}}$, while for the other parametrizations the order in which the particles appear is the same, but shifted to higher densities: $0.319\,\mathrm{fm^{-3}}$ for $\zeta=0.059$, $0.333\,\mathrm{fm^{-3}}$ for $\zeta=0.085$ and $0.349\,\mathrm{fm^{-3}}$ for $\zeta=0.129$. In other words, for a given central density, presented in Tables \ref{stars_models} and \ref{starsH_models}, one can track the particles population from Figure \ref{pop_l0040} to find out which degrees of freedom are present inside the star.

It was already shown in the previous section that strong magnetic fields decrease the central density of stars. 
%As one can infer from Table \ref{Hmodels}, for some magnetic fields distributions, the entral densities are lower than the threshold for the appearance of hyperons. 
From a microscopic point of view, it is also important to emphasize that Landau quantization has the effect of shifting the threshold of particle appearance \cite{Gomes:2014dka}. Nevertheless, the latter is a small effect compared to the former and it was, therefore, not taken into account in the calculations presented in this section. A thorough analysis of the former effect can be found for the MBF model in Ref. \cite{Dexheimer:2012qk} and in Ref. \cite{Gomes:2014dka}. 

\begin{figure}[!ht]
\includegraphics[width=0.85\linewidth]{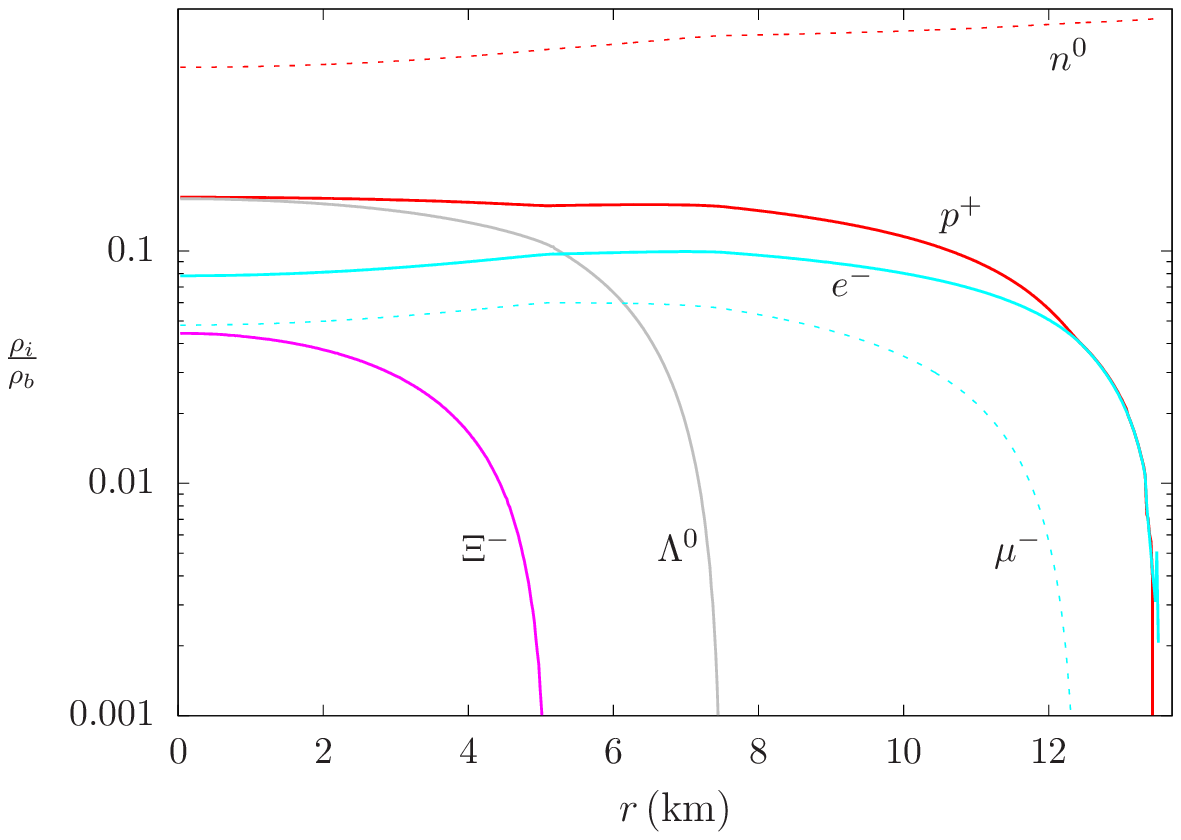} 
\includegraphics[width=0.85\linewidth]{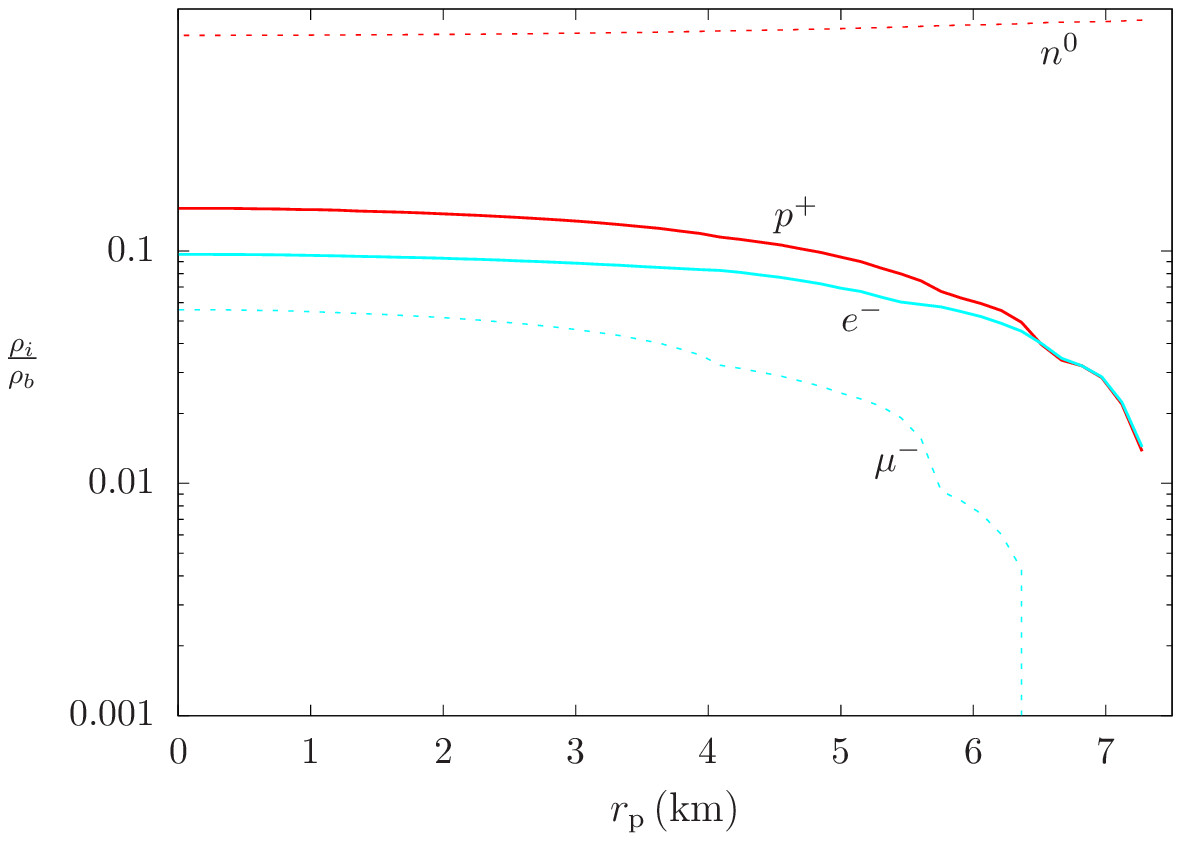} 

\caption{Particles population of the $M_b = 2.2\,M_{\odot}$ star for non-magnetic stars (top panel) and for magnetized stars with $j_0 = 3.5\times10^{15}\,\mathrm{A/m^2}$ (bottom panel), for the parametrization $\zeta=0.040$ of the MBF model. The vertical axis is the particle fraction normalized by the baryon density, and the horizontal axis is the (polar) radius. On the top panel, hyperons appear at $r_Y = 7.50\,\mathrm{km}$, and the total radius is $r_T = 14.03\,\mathrm{km}$. On the bottom panel, hyperons never appear ($r_Y = 0$) and the total polar radius is $r_p = 7.27\, \mathrm{km}$. }
\label{popraio}
\end{figure}

We now focus only on the two parametrizations which can describe massive hyperon stars, $\zeta=0.040,\,0.059$. 
From Table \ref{starsH_models} one can see that for $\zeta=0.040$, the highest magnetic field configuration $j_0=3.5\times10^{15}\mathrm{A/m^2}$ generates a $M_b=2.2\,\mathrm{M_{\odot}}$ star with a central density of $0.299\,\mathrm{fm^{-3}}$, which is lower than the $\Lambda$ threshold. This means that, for such a choice of parameters, the star will be entirely nucleonic.  The dramatic change in this specific star population is illustrated in Figure \ref{popraio}, which shows the particles population as a function of the radius for a non-magnetic star (top panel) and as a function of the polar radius for the highest magnetic field configuration (bottom panel), when the hyperon population vanishes entirely. Analogous results are found for the equatorial direction.

For the second highest magnetic field configuration $j_0=3.0\times10^{15}\mathrm{A/m^2}$, $\Lambda$ hyperons are allowed in a very small density interval of $0.064\,\mathrm{fm^{-3}}$ (difference between the central density and the $\Lambda$ hyperon threshold for this parametrization). 
The last column in Table \ref{starsH_models} shows the stellar radii fraction that contain hyperons in the equatorial direction $(r_Y/r_T)_{eq}$, which in this case is larger than $0.5$ ($50\%$ of the star). This is the case because the baryon density increases slowly towards the center of magnetic stars described by stiff EoS's. This behaviour can be seen in Figure \ref{rho_perfil}) in the polar direction, but it is even more drastic in the equatorial direction. In particular, $\Lambda$ hyperons reach the central fraction  $\rho_{\Lambda}/\rho_b = 0.07$ for  $j_0=3.0\times10^{15}\mathrm{A/m^2}$, and $\rho_{\Lambda}/\rho_b = 0.142$ for $j_0=2.0\times10^{15}\mathrm{A/m^2}$ for $\zeta=0.040$.
In all magnetic field configurations analyzed for this parametrization, only the case of $j_0=2\times10^{15}\,\mathrm{A/m^2}$ allows for the appearance  of $\Xi^-$ particles with a density threshold $\rho_b=0.377\,\mathrm{fm^{-3}}$. 
The largest interval of densities in which $\Lambda$ particles appear has a size of $0.127\,\mathrm{fm^{-3}}$ for the same value of the electric current amplitude $j_0$. Summarizing, it is necessary to have a surface and central magnetic fields higher than $B_s=3.8\times10^{17}\,\mathrm{G}$ and $B_c=8.01\times10^{17}\,\mathrm{G}$ for the total disappearance of hyperons (all assuming $\zeta= 0.040$).
% pop B= 0 

Still regarding the hyperon population content as a function of the radius \textcolor{red}{$(r_Y/r_T)_{eq}$}, one can also see from Table \ref{starsH_models} that this quantity increases when comparing the non-magnetic case and the case of $j_0=2\times10^{15}\,\mathrm{A/m^2}$, but then decreases for higher values of electric current amplitude. Such behavior comes both from the central density decrease and the increase of the stars' radius towards the equatorial direction. From Figure \ref{pop_l0040} one can see that the decrease of the central density from $\rho_b = 0.452\,\mathrm{fm}^{-3}$ for $j_0=0$ to $\rho_b = 0.415\,\mathrm{fm}^{-3}$ for $j_0=2.0\times10^{15}\mathrm{A/m^2}$ does not affect significantly the hyperon fraction and, moreover, the number of hyperon particles ($\Lambda$, $\Xi^-$) is the same, whereas the (equatorial) radius varies from $r_Y^{eq}=7.50\,\mathrm{km}$ to $r_Y^{eq}=7.65\,\mathrm{km}$, respectively. This means that the star gets larger in the equatorial direction, keeping roughly the same fraction of hyperons and, consequently, increasing the values of $(r_Y/r_T)_{eq}$. However, when higher magnetic field configurations are introduced, the central density decreases, not reaching the $\Xi^-$ hyperon threshold, lowering substantially the hyperon fraction and the $(r_Y/r_T)_{eq}$ value fraction.

Finally, we can also look at the $\zeta=0.059$ parametrization in Table \ref{starsH_models}. As this parametrization has an EoS softer than the previous one, higher central densities are reached and at least a small fraction of hyperons is present for all magnetic field configurations analyzed. For this parametrization, higher densities are reached (compared to $\zeta = 0.040$) for higher magnetic field configurations and, in particular, for $j_0=3.0\times10^{15}\mathrm{A/m^2}$ both $\Lambda$ and $\Xi^-$ hyperon thresholds are reached ($\rho_{b,\Lambda}=0.319\,\mathrm{fm^{-3}}$, $\rho_{b,\Xi^-}=0.040\,\mathrm{fm^{-3}}$). 
The  $j_0=2.0\times10^{15}\mathrm{A/m^2}$ case generates a larger interval of densities populated by hyperons  $0.261\,\mathrm{fm^{-3}}$.
The percentage of the stars populated by hyperons is also larger than for the stiffer EoS parametrization, reaching $\sim 65\%$ of the lowest magnetic field configuration  star in this analysis. For this parametrization, $\Lambda$ hyperons reach the central fraction $\rho_{\Lambda}/\rho_b = 0.09$ for  $j_0=3.5\times10^{15}\mathrm{A/m^2}$, $\rho_{\Lambda}/\rho_b = 0.167$ for  $j_0=3.0\times10^{15}\mathrm{A/m^2}$, and $\rho_{\Lambda}/\rho_b = 0.218$ for $j_0=2.0\times10^{15}\mathrm{A/m^2}$.
However, we stress that such a density interval containing hyperons  is much smaller than the one covered by non-magnetic neutron stars described by the MBF model, which is more than $\sim 0.5\,\mathrm{fm^{-3}}$ \cite{Gomes:2014aka}.

\begin{table*}[t]
  \caption{\label{starsH_models} Same as Table \ref{stars_models} for a $M_b=2.2\,\mathrm{M_{\odot}}$ star but now containing hyperons. The table includes a new column with the fraction of the star equatorial radius which contains hyperons (where $r_Y$ is the radius in which the hyperon population appears and $r_T$ is the radius at the surface).}
\begin{center}
\begin{tabular}{ccccccc}
 \hline
 
$\zeta$ & $j_0\,(10^{15}\mathrm{A/m^2})$ &  $\mu\,(10^{32}\mathrm{Am^2})$ & $ B_s\,(10^{17}\mathrm{G}) $ & $ B_c\,(10^{17}\mathrm{G}) $ & $\rho_c \,(\mathrm{fm^{-3}})$ & $(r_Y/r_T)_{eq}$  \\
  \hline \hline
    %&  &  &  &  \\
% 0.040 & $1.0$ & $0.57$  & $1.03$  & $3.48$ & 0.375 \tabularnewline
 0.040 & $0$ & n.a.  & n.a.  & n.a. & 0.452 & 0.534   \tabularnewline   
 0.040 & $2.0$ & $1.17$  & $1.45$  & $4.77$ & 0.415 & 0.542  \tabularnewline
 0.040 & $3.0$ & $2.06$  & $2.67$  & $7.08$ & 0.353 & 0.507 \tabularnewline
 0.040 & $3.5$ & $2.88$  & $3.78$  & $8.09$ & 0.299 & 0 \tabularnewline

\hline

% 0.059 & $1.0$ & $0.52$  & $0.67$  & $2.44$ & 0.432 \tabularnewline
 0.059 & $0$ & n.a.  & n.a.  & n.a. & 0.649& 0.631   \tabularnewline   
 0.059 & $2.0$ & $0.95$  & $1.45$  & $5.40$ & 0.589 & 0.647 \tabularnewline
 0.059 & $3.0$ & $1.70$  & $2.57$  & $7.77$ & 0.485 & 0.629 \tabularnewline
 0.059 & $3.5$ & $2.38$  & $3.53$  & $8.65$ & 0.385 & 0.597 \tabularnewline
  \hline\hline
  
  \end{tabular}
\end{center}
\end{table*}

% Bc and rhoc
%\begin{figure}[!ht]  %colocar o gráfico aqui ou no topo da página
%  \centering
%  \includegraphics[width=1.\linewidth]{bc_rhoc_paper.eps}
%  \caption{Central density dependence on the central magnetic field, for the $M_b=2.2\,\mathrm{M_{\odot}}$ star, for the parametrizations $\zeta=0.040$ and $\zeta=0.059$.}
%\label{bc_rhoc}
%\end{figure}

\section{Final discussion}

We have investigated the impact of different parametrizations of the many-body forces (MBF) model with different degree of stiffness on the properties of highly magnetized nucleon and hyperon stars. 
Four parametrizations of the model were used, which are all able to reproduce nuclear properties at saturation. We calculated global properties of neutron stars self-consistently by including Landau quantization effects on the EoS and solving the Einstein-Maxwell's field equations for a poloidal magnetic field distribution in an axi-symmetric metric (2D solutions).

First, we confirmed that microscopic magnetic field effects on the equation of state are not significant for the description of the global properties of nucleonic stars using our EoS, in agreement with previous works that had found similar results for quark and hybrid stars  \cite{Chatterjee:2014qsa,Franzon:2015sya}. A comparison for magnetic neutron stars done by solving the TOV equations (including the pure magnetic field contribution isotropically) and the 2D self-consistent calculations was carried out for the first time. Our results show that neglecting the deformation of  stars %as well as including the magnetic field correctly in the structure calculations 
leads to an overestimate of more than $12\%$ for both gravitational and baryon maximum masses, and an 
underestimate of more than $20\%$ for the equatorial radius of  $1.4\,\mathrm{M_{\odot}}$ stars. From these results, we concluded that using the TOV equations to describe magnetic neutron stars is not the correct approach, since it generates large errors in the calculation of global properties of stars already for central magnetic fields lower than $\sim 10^{18}\,\mathrm{G}$.

In order to focus on the effects of many-body forces on magnetic stars, the baryon mass of stars was fixed to $2.2\,\mathrm{M_{\odot}}$ and we varied both the magnetic field configuration (through the electric current amplitude $j_0$) and the many-body forces parameter $\zeta$, responsible for the different stiffness of the EoS. The many-body forces parameter has the effect of shielding the scalar couplings of the model, enhancing the repulsion among particles for low values of $\zeta$. As a consequence, more massive, as well as larger stars are generated for the stiffest parametrization of the MBF model. These results hold for both non-magnetic and magnetic stars.

In addition, applying different parametrizations of the model to describe highly magnetic stars, it was shown that the softer parametrizations allow for higher central densities and, consequently, higher central magnetic fields. Also, because of the larger radii described by stiff parametrizations of the model, these stars are less compact and, hence more easily deformed. It is important to emphasize that, because of the choice of poloidal magnetic field distributions, oblate shapes are favored for highly magnetic stars. Note however that, several studies suggest that toroidal contributions might play an important role on the stability of magnetic stars \cite{Braithwaite:2005ps,Marchant:2010yj,Lasky:2011un,Ciolfi:2013dta,Akgun:2013aq,Mitchell21022015,Armaza2015,Mastrano11032015}. Still, even in this case we expect our qualitative results to hold. %Such configurations make the stars more prolate and it is possible that the combination of both toroidal and poloidal contributions would lead to less deformed stars. 

It was also shown in this work that strong magnetic field distributions decrease the  central densities of neutron stars, which has a very large impact on the particle population of these objects.
In particular, for the parametrizations in agreement with the observational data for non-magnetic hyperon stars ($\zeta=0.040.\,0.059$), hyperons populate only a small interval of densities, but a significant portion of the stellar volume. 
As already shown in Ref. \cite{Franzon:2016iai} for one EoS, the strangeness is overall lower in a star with larger magnetic field, but it stays steady for a larger portion of its radius. 

%In this work we have investigated for the first time the impact of different degrees of stiffness of the nuclear EoS in magnetic neutrons stars. 
Although we have used a specific nuclear model to describe neutron stars in this work, our results are general and can (and should) be tested with different models. Nevertheless, it would be interesting to compare them quantitatively with results from other models. 
As a future perspective of this work, studies of the thermal evolution of such objects will be important for the search of potential observational signals of  hyperon stars and their creation during the decay of stellar magnetic field over time. Finally, studies including both strong magnetic fields and fast rotation are important to describe the initial stages of the lives of highly magnetized neutron stars and are already been carried out.

%\newpage
\begin{acknowledgements}
The authors acknowledge the support from NewCompstar, COST Action MP 1304.
R.O. Gomes and S. Schramm acknowledge support from HIC for FAIR.
This work was also partially supported by grant Nr. BEX 14116/13-8 of the PDSE CAPES and Science without Borders programs which are an initiative of the Brazilian Government.

\end{acknowledgements}

%\bibstyle{apsrev4-1}  
\bibliographystyle{aa} % style aa.bst
\bibliography{magnetic_bib}

\end{document}